\documentclass[amsmath,amssymb,aps,prb,twocolumn]{revtex4-2}
\usepackage[utf8]{inputenc}
\usepackage{graphicx}
\usepackage{subcaption}
\usepackage{dcolumn}
\usepackage{bm}
\usepackage{hyperref}
\usepackage{amsmath}
\usepackage{amsfonts}
\usepackage{amssymb}
\usepackage{natbib}
\usepackage{textcomp,gensymb}
\usepackage{color}
\usepackage{graphicx} 
\usepackage{caption}
\begin{document}

\title{Temperature anomaly of the $V_{Si}$ and $V_{C}$ vacancy spin coherence time in $4H$-SiC}
\author{P. Chrostoski$^{\dag,\ddag}$}
\author{Ifeanyi Innocent Onwosi$^{\ast}$}
\author{D. H. Santamore$^{\ast}$}
\affiliation{$^{\dag}$Sandia National Laboratories, Livermore, California 94550, USA}
\affiliation{$^{\ddag}$Department of Physics, Waldorf University, Forest City, IA, 50438, USA}
\affiliation{$^{\ast}$Department of Physics and Engineering Physics, Delaware State University, Dover, DE 19901, USA}

\begin{abstract}
Increasing the spin coherence time ($T_{2}$) is a major area of interest for spin defect systems such as the silicon ($V_{Si}$) and carbon ($V^{\pm}_{C}$) vacancies in $4H-$SiC. Usually as temperature increases, $T_{2}$ decreases due to the thermal bath. Observations of electron-paramagnetic resonance and direct systematic measurements of $T_{2}$ has seen an anomaly where $T_{2}$ increases with increasing temperature. In this work, we investigate the mechanisms that cause the $T_{2}$ temperature anomaly. We find that due to a spontaneous symmetry lowering from a motional Jahn-Teller distortion, a polaron quasiparticle is generated from the vibronic coupling. Initially, for temperatures from $8$ to $20-40$K, the coherence temperature dependence is dominated by phonon-assisted spin relaxation. At temperatures around $20-40$K, depending on the vacancy, a thermally activated polaron hopping turns on and motional narrowing dominates and increases $T_{2}$ with increasing temperature. As temperatures reach $120-160$K, the energy barrier gets high enough to slow the polaron hopping. At this point the Larmor precession dominates, leading to decoherence. Our calculated temperature-dependent coherence agrees with what has been seen experimentally, giving a full theoretical framework for the mechanisms that cause the $T_{2}$ temperature anomaly of increasing $T_{2}$ with increasing temperature. The theoretical framework presented here also gives insight into these mechanisms being a probable universal phenomenon that could occur in many other defect center spin systems.
\end{abstract}
\date{\today }
\maketitle

\section{Introduction}

Localized defects in solids have shown a potential to be useful in applications such as single photon emission \cite{CJI14,LIB15}, sensing \cite{KSF14}, and quantum computing \cite{EMK95}. Two such defects have been of particular interest; the nitrogen-vacancy center in diamond (NV center) and the carbon vacancy ($V^{\pm}_{C}$), silicon vacancy ($V_{Si}$), and divacancies ($V^{\pm}_{C}$-$V_{Si}$) in silicon carbide (SiC). Spin states in NV centers can be manipulated and controlled through the use of pulse sequences (such as dynamical decoupling) yielding coherence times on the order of milliseconds at room temperature and hundreds of milliseconds at 77$\mathrm{K}$ \cite{BPJ13}. The issue is that the manufacturing of these diamonds and the integration into applications can be difficult and expensive. SiC is a material that has similar quantum mechanical properties to NV centers but potentially easier fabrication \cite{BBS11}.

There are many different polytypes of SiC which can have a variety of defects with different spin properties \cite{RFK12,SSB12,SCH06}. The unique spin properties of the individual defects (such as the $V_{Si}$ defect) can be optically initialized, addressed, and read out through optically detected magnetic resonance (ODMR) \cite{CSD15}. The $4H$-SiC polytype, for instance, is made up of two inequivalent lattice sites; the hexagonal ($h$) and quasicubic ($k$) where the silicon vacancy can exist. These different sites can have a variety of coherence times ($T_{2}$). Coherence lifetimes on the order of 100$\mu s$ have been recorded for the $V_{Si}$ defect \cite{MLR15}. On the other hand, the spin coherence lifetimes of the $V^{\pm}_{C}$-$V_{Si}$ defects have been seen to be in the range of tens of microseconds to a few milliseconds \cite{CFA15,FBC13,KBH11}. Similarly to NV center-based spin systems, these coherence times can be extended through the use of dynamical decoupling techniques \cite{KBH11}. 

It was not until future work that systematic measurements of the coherence temperature dependence were obtained \cite{FKI15,ECM17}. From these works looking at $V_{Si}$ in $4H$-SiC and the divacancies in $6H$-SiC, it was seen that as the temperature increased to $40$K, the coherence time diverted from the expected decrease with increasing temperature due to phonon-assisted spin relaxation \cite{SKS17}. The coherence time eventually peaked around $150-160$K before returning to the expected trend (decreasing $T_{2}$ with increasing temperature). This unexpected increase in coherence has been attributed to a motional Jahn-Teller (JT) distortion \cite{ECM17,UIM04,UII05,BMO01}. Until now, there has been no theoretical analysis to understand what mechanisms that arise from this motional JT distortion, causing the coherence time anomaly in SiC vacancy center spin systems.

In this work, we model how motional JT distortion leads to the coherence time temperature anomaly. We propose two mechanisms: lifetime broadening arising from phonon-assisted spin relaxation, and motional narrowing arising from thermally activated reorientation. In defect center based spin systems, the fluorescence lifetime broadening is an unfortunate side effect due to non-idealities that cause decoherence of the spin state. On the other hand, in inhomogeneous systems (such as defect centers in $4H$-SiC), electron and nuclear motion can cause the spectral linewidth to narrow. The narrowing of the spectral linewidth then directly implies an increase in coherence time. This is the motional narrowing effect that has been observed in organic materials \cite{TKY06,MAG11,MHT08,MMH10,MKT12} and theoretically studied \cite{RK54,KT54,PWA54}. Motional narrowing has been known for decades \cite{CS90,NB48}, but understanding its effects and how it arises in different systems is still of interest in understanding the mechanisms causing inhomogeneous broadening in experiment \cite{SM19}. One case where motional narrowing is prominent is in systems that generate a polaron quasiparticle. Polarons form due to phononic couplings similar to motional JT distortions arising due to vibronic couplings. 

In Sec. \ref{T2JT} we describe the spin Hamiltonian for the vacancy center and describe the coherence lifetime contribution due to motional JT distortion using Redfield relaxation theory. Redfield relaxation theory describes the coherence time effect due to magnetic field fluctuations. In Sec. \ref{BfieldTempDepend} we model the $x-$, $y-$, and $z-$ coordinate magnetic field temperature dependent fluctuations. The $z-$axis contribution is modeled through the nuclear magnetic field of the surrounding silicon and carbon atoms. Now when considering the paramagnetic nature of the defect center, we can model the $x-$ and $y-$component of the magnetic field fluctuations with the Brillouin function. In Sec. \ref{LifetimeTempDepend} we link the vibronic couplings of the JT distortion to the generation of a polaron. We then model the characteristic lifetime via the thermally activated Holstein two-site polaron hopping rate. The last piece to model is the electron energy reduction due to the JT distortion. This requires understanding the vibronic couplings. In Sec. \ref{vibroniccoupling} we analytically model the linear vibronic coupling constant ($F$) and atomistically calculate the elastic force ($K$) and quadratic coupling ($G$) constant for the $V^{\pm}_{C}$ and $V_{Si}$ vacancies which are necessary to model the potential energy surface (PES) of the JT distortion. Lastly, we bring these all together to be able to predict the coherence time temperature dependence for the $V^{\pm}_{C}$ and $V_{Si}$ vacancy center defects of $4H-$SiC.

The work presented here is able to calculate the coupling constants necessary to determine the adiabatic potential energy surface (APES) for the $V^{\pm}_{C}$ and $V_{Si}$ vacancies. Our analytical and atomistic methods are able to calculate $F$, $K$, and $G$. With these values, we are able to generate the APES and calculate the Jahn-Teller stabilization energy ($E_{JT}$) and energy barrier ($\delta_{JT}$) for each vacancy center. The calculated values of $E_{JT}$ and $\delta_{JT}$ fall within the expected value range compared to the values calculated for the divacancies in $4H-$SiC. From here we build a theoretical model that is able to reproduce the observations of an increase in $T_{2}$ with increasing temperature for the $V^{\pm}_{C}$ and $V_{Si}$ vacancies. The temperature ranges where the motional narrowing turns on and then off and the predicted $T_{2}$ times are in good agreement with what is seen in the experiments done by Embley \textit{et al} \cite{ECM17} and Umeda \textit{et al} \cite{UIM04,UII05}. Beyond this, our theoretical model gives insight into a probable universal phenomenon of polaron formation leading to thermally activated motional narrowing for defect center spin systems.

\section{Model}
\subsection{$T^{*}_{2}$ due to Jahn-Teller distortion} \label{T2JT}
To study the coherence time impact from the motional JT distortion of the vacancy, we start by considering the spin Hamiltonian plus a time-varying perturbation arising from the motional JT distortion
\begin{align}
\hat{H} = \hat{H}_{0} + \hat{H}_{1}(t).
\end{align}
Here $\hat{H}_{0}$ is the spin Hamiltonian and $\hat{H}_{1}(t)$ is the perturbation. The spin Hamiltonian is given by \cite{UIM04}
\begin{align}
\hat{H}_{0} \nonumber &= \mu_{B}\textbf{B} \cdot \textbf{g} \cdot \textbf{S} \\
&+ \sum_{n}\left[ \textbf{S} \cdot \textbf{A}(\mathrm{Si}_{n}) \cdot \textbf{I}(\mathrm{Si}_{n}) - \mu_{N}g_{N}\textbf{I}(\mathrm{Si}_{n}) \cdot \textbf{B}\right].
\end{align}
Here $\mu_{B,N}$ are the Bohr and nuclear magnetons respectively, $\textbf{B}$ is the magnetic field, $\textbf{g}$ is the g tensor, $g_{N}$ is the nuclear g factor, $\textbf{S}$ is the electron spin operator, $\textbf{A}(\mathrm{Si}_{n})$ is the hyperfine (HF) tensor for the $n_{th}$ Si atom, and $\textbf{I}(\mathrm{Si}_{n})$ is the nuclear spin operator. The perturbation piece, $\hat{H}_{1}(t)$, can be written with the following form \cite{AGR57}
\begin{align}
\hat{H}_{1}(t) = \sum_{n}F_{n}(t)\hat{A}_{n},
\end{align}
where $F_{n}(t)$ are random functions of time and $\hat{A}_{n}$ are the eigenoperators. In this picture of spin dynamics, the perturbation due to the spin-lattice interaction can be modeled using the Redfield master equation \cite{CS90,AA61,AGR57}. Redfield relaxation theory describes the spin dynamics of a system in terms of stochastic fluctuations of an effective magnetic field and a characteristic frequency \cite{CS90,AA61,AGR57,SCL19}. 

Let $T^{JT}_{2}$ denote the coherence lifetime contribution from the motional JT distortion described by the Redfield relaxation \cite{CS90,AA61}
\begin{align}
\frac{1}{T^{JT}_{2}} = \frac{1}{T_{2}} + \frac{1}{2T_{1}}. \label{T2total}
\end{align}
Here the spin-lattice relaxation time, $T_{1}$, is related to the magnetic field fluctuations in the x- and y-directions by \cite{CS90,SCL19}
\begin{align}
\frac{1}{T_{1}} = \gamma_{e}^{2}\left( \delta B_{x}^{2}+ \delta B_{y}^{2}\right)\frac{\tau_{c}}{1+\omega_{0}^{2}\tau_{c}^2}.
\label{T1effect}
\end{align}
Here $\delta B_{x,y}$ are the magnetic field fluctuation amplitudes in the x- and y-directions, $\gamma_{e}$ is the electron gyromagnetic ratio, $\omega_{0}$ is the Larmor frequency, and $\tau_{c}$ is the characteristic lifetime. We can then write the coherence time, $T_{2}$, contribution in terms of the magnetic field fluctuations in the z-direction
\begin{align}
\frac{1}{T_{2}} = \gamma_{e}^{2} \delta B_{z}^{2}\tau_{c}, \label{T2effect}
\end{align}
where $\delta B_{z}$ is the magnetic field fluctuation amplitude in the z-direction. Substituting equations (\ref{T1effect}) and (\ref{T2effect}) into equation (\ref{T2total}) we obtain \cite{CS90,SCL19},
\begin{align}
\frac{1}{T^{JT}_{2}} = \gamma_{e}^{2}\delta B_{z}^{2}\tau_{c} + \frac{1}{2}\gamma_{e}^{2}\left(\delta B_{x}^{2}+\delta B_{y}^{2}\right)\frac{\tau_{c}}{1+\omega_{0}^{2}\tau_{c}^2}. \label{T2Complete}
\end{align}

We label the coherence time contribution due to motional JT distortion as $T^{JT}_{2}$ due to the fact that it is a single source that affects the total coherence time of a spin system which is defined by the inhomogeneous dephasing time $T^{*}_{2}$. This inhomogenous dephasing time is \cite{BSB20,CKS22}
\begin{align}
\frac{1}{T^{*}_{2}} \approx \frac{1}{T^{elect}_{2}} + \frac{1}{T^{mag}_{2}} + \frac{1}{2T_{1}},
\end{align}
where $T^{elect}_{2}$ encompasses coherence times predicted by electric field fluctuations, $T^{mag}_{2}$ encompasses the magnetic field fluctuations on coherence, and as above, $T_{1}$ is the spin-lattice relaxation. This longitudinal relaxation can come about from sources such as phonon-assisted spin-relaxation and couplings to dielectric materials according to Fermi's golden rule \cite{CKS22,SKS17}. Thus, the motional JT distortion will contribute to both $T^{mag}_{2}$ and $T_{1}$. 

\subsection{Temperature dependent magnetic fields of $T^{JT}_{2}$} \label{BfieldTempDepend}
Thus far, we have not considered the temperature dependence of $T^{JT}_{2}$. It follows from equation (\ref{T2Complete}) that the temperature dependent fluctuations of the components of $\textbf{B}$ and $\tau_{c}$ account for the temperature dependence of $T^{JT}_{2}$. To start, we need to understand how the JT distortion is changing the vacancy center. Previous work by Umeda \textit{et al} \cite{UIM04,UII05} looking at $V^{\pm}_{C}$ attributed the narrowing of the $\mathrm{Si}_{1}$ hyperfine line in their electron paramagnetic resonance measurement to a thermally activated reorientation of the defect. As the temperature of the system increases, this reorientation causes the unpaired electron to begin jumping (or hopping) between different electronic states. The electron will hop between the following bondings: Si$_{1}-$Si$_{2}$, Si$_{1}-$Si$_{3}$, and Si$_{1}-$Si$_{4}$ for the carbon vacancy or C$_{1}-$C$_{2}$, C$_{1}-$C$_{3}$, C$_{1}-$C$_{4}$ for the silicon vacancy. This hopping between the electronic states will give rise to magnetic field fluctuations in the $x-$, $y-$, and $z-$ directions. 

With this in mind, we can look at the magnetic field fluctuations by considering what we know about SiC and how the vacancies change the magnetic properties. Pure SiC is a non-magnetic semiconductor. However, the vacancy defect centers are paramagnetic. Due to this paramagnetic nature, the magnetic field of the vacancy center will be temperature dependent. For a classical system, one can model the temperature dependence of a paramagnetic using Curie's magnetic field. As we are considering a quantum mechanical system, it is more accurate to consider the Brillouin function. Like Curie's magnetic field, the Brillouin function will describe the magnetic field as a function of temperature, but also considers the total angular momentum of a quantum system (more importantly, the spin). As derived in appendix (\ref{Appdx:paramagnetism}), the Brillouin function is given by
\begin{align}
B_{J}(C) = \frac{2J+1}{2J}\coth\left(\frac{2J+1}{2J}C \right) - \frac{1}{2J}\coth\left(\frac{1}{2J}C \right), \label{BrillBfield}
\end{align}
where the constant $C = Jg\mu_{B}B/k_{B}T$ is the ratio of the Zeeman energy of the magnetic moment to the thermal energy. $J$ is the total angular momentum, $k_{B}$ is the Boltzmann constant, $T$ is temperature, and $B$ is an external magnetic field. Now, we note here that the $V_{C}^{\pm}$ are spin$-1/2$ systems. The Brillouin function of the magnetic field can be simplified for a spin-$1/2$ system when plugging in $J=1/2$ to obtain
\begin{align}
B_{J} = \tanh \left( \frac{Jg\mu_{B}B}{k_{B}T}\right).
\end{align}
We are also concerned with the $V_{Si}$ vacancy centers in $4H$-SiC, which are spin$-3/2$ ($J=3/2$) systems. With that, we can use the simplified form of the Brillouin function (see appendix (\ref{Appdx:paramagnetism})),
\begin{align}
B_{J} = \left( \frac{J+1}{3}\right)\frac{g\mu_{B}B}{k_{B}T}. \label{reducedBrillBField}
\end{align}

As the electron hops from site to site, it will experience the nuclear magnetic field from the new site, allowing us to describe the external magnetic field in equation (\ref{reducedBrillBField}) as the nuclear magnetic field. Using magnetic moment data for the vacancy center in $4H$-SiC from Kukushinkin \textit{et al} \cite{KO22} and considering that the nuclei of each site is a magnetic dipole (see appendix (\ref{Appdx:nuclearmagfield}), we get the following nuclear magnetic field
\begin{align}
B_{N} = \frac{3\mu_{0}\mu_{B}}{2\pi r^{3}}.
\end{align}
Here $\mu_{0}$ is the magnetic permeability of free space and $r$ is the atomic radius.

Another temperature dependent fluctuation of the magnetic field we need to take into account will come from the electron hopping. As the electron hops to a new nuclear site with some momentum, it will experience a Lorentz force due to the new magnetic environment which can be described by
\begin{align}
\frac{\partial \textbf{p}}{\partial t} = q\left(\textbf{v}_{q} \times \textbf{B}\right),
\end{align}
where $\textbf{p}$ is the momentum and $\textbf{v}_{q}$ is the velocity of the charge. We neglect the electric field Lorentz force as we are currently only worried about the magnetic field fluctuations of the local environment of the electron spin as stated above. We get the magnetic field to be (see appendix \ref{append:LarmorBField})
\begin{equation}
B_{L}(T) = \left(\frac{2}{qc^{2}r}\right)v_{q}(T)E_{q}(T), \label{LarmMagField}
\end{equation}
We label this magnetic field $B_{L}$ as this magnetic field of the external field will cause the electron spin to experience Larmor precession. Here $v_{q}(T)$ is the temperature dependent charge velocity, $E_{q}(T)$ is the temperature dependent electron energy in electron volts, $q$ is the electron charge, and $c$ is the speed of light. The temperature dependence of the charge velocity as it hops comes from equipartition. Whereas the temperature dependence of $E_{q}$ comes from being effected by the JT-distortion through the JT energy ($E_{JT}$). The temperature dependence of $E_{JT}$ arises from the anharmonicity of the motional JT distortion. We then write $E_{q}(T)$ as
\begin{align}
\Delta E_{q}(T) = E_{q}\left(1 \pm \frac{E_{JT}(T)}{E_{q}} \right).
\end{align}
Determining the temperature dependence of $E_{JT}$ will require characterizing the adiabatic potential energy surface (APES) and performing a full temperature dependent calculation of important coupling constants described in section (\ref{vibronics}). 

We are now able to model the magnetic field fluctuations along each axis. The magnetic field due to the nuclei at each site describes the z-axis fluctuations. We thus can describe $\delta B_{z}$ with Eqn. (\ref{reducedBrillBField}). Whereas, the Lorentz force will give Larmor precession fluctuations along the x- and y-axis allowing us to describe $\delta B_{x,y}$ with Eqn. (\ref{LarmMagField}). With the magnetic field temperature dependence determined, we now need to understand the temperature dependence of the polaron hopping characteristic lifetime ($\tau_{c}$).

\subsection{Characteristic lifetime of polaron hopping, $\tau_{c}$} \label{LifetimeTempDepend}
To be able to determine the temperature dependent $\tau_{c}$, we need to look deeper into the motional JT distortion. Recent work on the effects of temperature on the silicon vacancy in $4H$-SiC shows a polaronic gap for the vacancy excited states \cite{UTM20}. The polaronic, as opposed to electronic, excited state comes from the orbital degeneracy leaving the defect center vulnerable to the symmetry-lowering JT distortion \cite{UTM20,IDS17,AG12,IBB06}. The strong coupling between the electron and phonons generates the polaron \cite{LP08}. Due to this polaronic nature and the fact that the hopping will only occur between two sites (i.e., between the sites Si$_{1}-$Si$_{2}$, Si$_{1}-$Si$_{3}$, Si$_{1}-$Si$_{4}$ for the carbon vacancy or C$_{1}-$C$_{2}$, C$_{1}-$C$_{3}$, C$_{1}-$C$_{4}$ for the silicon vacancy), we can describe the hopping with Holstein's two-site hopping model \cite{TH781,TH782}.

To determine the two-site hopping rate, we consider the probability of an electronic transition between site $p$ (Si$_{1}$ for the carbon vacancy and C$_{1}$ for the silicon vacancy) and $p+1$ (Si$_{2,3,4}$ for the carbon vacancy and C$_{2,3,4}$ for the silicon vacancy). We can then consider the vibrational electron-phonon coupling of the two sites in a molecular crystal model described by the basic molecular Hamiltonian and the Holstein Hamiltonian \cite{TH59,TH781,TH782,YC14,FT07}
\begin{equation}
\hat{H} = \hat{H}_{L} + \hat{H}_{e} + \hat{H}_{eL}. 
\end{equation}
Here $\hat{H}_{L}$ is the vibrational Hamiltonian, consisting of the kinetic and potential energies of vibration of an array of independent diatomic molecules that takes into account the coupling to nearest-neighbors,
\begin{equation}
\hat{H}_{L} = \sum_{n=1}^{N}\frac{-\hbar^{2}}{2M}\frac{\partial^{2}}{\partial r_{n}^{2}} + \frac{1}{2}M\omega_{0}^{2}r_{n}^{2} + \frac{1}{2}M\omega_{1}^{2}r_{n}r_{n+1}.
\end{equation}
Here $M$ is the mass, $r_{n,n+1}$ are the positions of site $n$ and the nearest-neighbor site $n+1$, and $\omega_{0,1}$ are the vibrational frequencies. $\hat{H}_{e}$ and $\hat{H}_{eL}$ are the electron transfer and electron interaction with the local vibrational mode Hamiltonians which make up the Holstein Hamiltonian such that $\hat{H}_{e}+\hat{H}_{eL}=\hat{H}_{Hol}$ and is described by \cite{TH59,TH781,TH782,YC14,FT07},
\begin{align}
\hat{H}_{Hol} = -A\sum_{i}\sum_{j}c_{i}^{\dagger}c_{i}+\hbar \nonumber & \omega_{i}\sum_{i}b_{i}^{\dagger}b_{i}\\
&-\lambda\sum_{i}\left( b_{i}^{\dagger}+b_{i}\right)c_{i}^{\dagger}c_{i}.
\end{align}
Here $A$ is the intersite hopping integral amplitude (that is, the nearest neighbor electron-transfer parameter), $\hbar$ is Planck's constant, $c_{i}^{\dagger}$ and $(c_{i})$ are the electron creation and annihilation operators, $b_{i}^{\dagger}(b_{i})$ are the phonon creation and annihilation operators, $\lambda$ is the electron-phonon coupling, and $\omega_{i}$ is the phonon angular frequency. The key parameter we need from the Holstein model is the two-site hopping which will describe the characteristic lifetime, $\tau_{c}$. Following the Holstein model, this two-site hopping rate can be described by \cite{TH781,TH782}
\begin{align}
\nu_{p \xrightarrow{} p+1} = \int \frac{\Phi(E_{a})}{2\pi\hbar Z_{0}} \mathrm{e}^{\frac{-E_{a}}{k_{B}T}} dE,
\end{align}
where $\Phi(E_{a})$ is a thermodynamic average of the electron transition wave function, and $Z_{0}$ is the vibrational partition function. This integration (see appendix \ref{append:HolstModel} and \ref{append:TwoSiteHopping}) will give the thermally activated two-site hopping due to the reorientation of the vacancy \cite{TH781,TH782},
\begin{equation}
\nu_{p \xrightarrow{} p+1} =\frac{1}{\tau_{c}} = \frac{\lambda^{2}}{2\hbar} \sqrt{\frac{\pi}{E_{a}k_{B}T}} e^{-\frac{E_{a}}{k_{B}T}},
\end{equation}
where $E_{a}$ is the thermal activation energy of the hopping.

\subsection{Jahn-Teller vibronic coupling and APES} \label{vibroniccoupling}
\subsubsection{Vibronic interactions} \label{vibronics}
Now that we have a way to describe the characteristic lifetime of the thermally activated reorientation, the last piece is to determine how to model the temperature dependence of $E_{JT}$. This will require the modeling of the vibronic coupling within the system. Vibronic interactions at their core are electron-nuclear interactions due to nuclear motions in both degenerate and non-degenerate states. The Hamiltonian for the vibronic interactions is generally given by \cite{IBB06}
\begin{align}
    H = H_{r} + H_{Q} + V(r,Q),
\end{align}
where $H_{r}$ is the electronic component which includes the kinetic energy of the electrons and the interelectronic electrostatic interaction, $H_{Q}$ is the kinetic energy of the nuclei, and $V(r,Q)$ is the interaction energy between the nuclei and electrons. From the potential energy surface, we can determine $E_{JT}$ \cite{IBB06,ASS11,ZSV24}. Due to this fact, we will ignore the kinetic energy terms and look only at the potential energy term $V(r,Q)$.

We can write $V(r,Q)$ as a series expansion of small displacements of the nuclei about the nuclei's origin $Q=0$. We describe the positions in symmetrized coordinates $Q$, instead of using the usual position vector ($\Vec{R}$) in Cartesian coordinates. Symmetrized units are commonly used to describe the components of unit cells of materials that can be off-axis in Cartesian coordinates. The series expansion of the potential function is \cite{IBB06},
\begin{align}
    V(r,Q) = V(r,0) + \nonumber &\sum_{\alpha} \left(\frac{\partial V}{\partial Q_{\alpha}}\right) Q_{\alpha}\\ 
    & + \frac{1}{2}\sum_{\alpha,\beta}\left( \frac{\partial^{2} V}{\partial Q_{\alpha} \partial Q_{\beta}} \right) Q_{\alpha}Q_{\beta} + \cdots,
\end{align}

This expansion shows that the wave function of the Schr\"{o}dinger equation will have two parts, a radial term dependent on $r$ and a term depending on the nuclear coordinates $Q$. Neglecting small nuclear displacements arising from nuclear dynamics or vibronic coupling amounts to adopting the Born-Oppenheimer approximation. Instead, we will consider there will be small displacements about $Q=0$ to determine specific coupling constants of the vibronic interactions (see appendix \ref{append:JTEffect}). These coupling constants will be key for determining $E_{JT}$. Following the text of Bersuker \cite{IBB06}, we get the the linear vibronic coupling ($F$), the quadratic vibronic coupling ($G$), and a force constant ($K$) to be
\begin{align}
    F &= \langle \Phi_{n}| \left( \frac{\partial V}{\partial Q_{\alpha}}\right) |\Phi_{n^{'}}\rangle, \nonumber \\
    G &= \langle \Phi_{n}| \left( \frac{\partial^{2} V}{\partial Q_{\alpha} \partial Q_{\beta}} \right)_{\alpha \neq \beta} |\Phi_{n^{'}}\rangle, \nonumber \\
    K &= \langle \Phi_{n}| \left( \frac{\partial^{2} V}{\partial Q_{\alpha} \partial Q_{\beta}} \right)_{\alpha = \beta} |\Phi_{n^{'}}\rangle.\label{coupling}
\end{align}
The importance of the coupling constants come in that they are needed to be able to calculate the APES which is needed to determine $E_{JT}$. The expression for the APES in symmetrized coordinates (see again appendix \ref{append:JTEffect}) can be written as \cite{IBB06}
\begin{align}
\epsilon(\rho,\phi) = \frac{1}{2}K\rho^{2} \pm \rho\left[ F^{2}+G^{2}\rho^{2}+2FG\rho \cos{3\phi}\right]^{1/2}, \label{APES}
\end{align}
where $\rho=\sqrt{(Q_{x}^{2}+Q_{y}^{2})}$ and $\phi=\arctan(Q_{y}/Q_{x})$. Equation (\ref{APES}) shows the dependence on different coupling constants: the linear vibronic coupling ($F$), the quadratic vibronic coupling ($G$), and a force constant ($K$).

These couplings have been determined for the NV center and the divacancy in $4H$-SiC through non-temperature-dependent density functional theory (DFT) calculations. To be able to study a temperature-dependent APES, we need to determine the temperature dependence for the coupling constants. The anharmonicity of the motional JT effect implies that these couplings will, in fact, be temperature dependent. In $4H$-SiC, the vacancy center symmetry is $C_{3V}$ which is trigonal symmetry \cite{UIM04,UII05}. The temperature dependence of the linear vibronic coupling  $F$ for a trigonal symmetry can be analytically determined as derived by Sarychev \textit{et al} \cite{SHB20}
\begin{align}
F = \sqrt{\frac{27\alpha(T)}{4k_{E_{0}}}\frac{c_{E_{0}}k_{B}T}{n_{V_{C,Si}} a_{0}^{2}}}. \label{Fanalytic}
\end{align}
Here $c_{E_{0}}$ is the elastic modulus, $n_{V_{C,Si}}$ is the concentration of vacancies, $a_{0}$ is the distance between atoms, $k_{E_{0}}$ is the magnitude of the vibrational wave number, and $\alpha(T)$ is the attenuation.

Analytical solutions for the quadratic coupling ($G$) and force ($K$) constants are not available. We instead need to look closely at the $G$ and $K$ terms of Eqn. (\ref{coupling}). These couplings are described in a way that is analogous to a method used for calculating the vibrational modes of phonon interactions in materials: the dynamical matrix \cite{YXZ21,ZDP22}.

\subsubsection{The dynamical matrix} \label{dynamicalmatrix}

Let us consider again the vibrational interactions between the electron and the nuclei. We can make the analogy that the electron and nuclei make up a two mass spring system. With that in mind, the second order spatial derivative around the equilibrium atomic positions will describe a force constant (i.e., dynamical) matrix $\textbf{R}_{i}$ \cite{BGC01}
\begin{align}
\Phi_{ij}^{ab} = \frac{1}{\sqrt{m_{a}m_{b}}}\frac{\partial^{2} U}{\partial r_{i,a}\partial r_{j,b}},
\end{align}
where $U$ is the potential energy surface, $r_{a,b}$ are the displacements of particle $a$ and $b$ respectively, $m_{a,b}$ are the masses of the particle $a$ and $b$, and the indices $i$ and $j$ represent spatial directions (i.e. the Cartesian direction the displacement is taking place). If we consider vibrational dynamics, the second derivative of the potential is a force constant. For this reason, the dynamical matrix is often referred to as the force constant matrix. In \cite{ASS11,ZSV24}, the force and quadratic coupling constants are extracted by fitting to potential energy curves of DFT calculations. There is a method to calculate this value more directly.

To do this, observe from Eqn. (\ref{coupling}) that the force constant is defined by the second spatial derivative of the potential function. This potential function can be described by an interatomic potential function which will describe the interaction between the atoms of the crystal structure. The use of interatomic potentials is a semi-empirical theoretical method that uses corrections from experiments to allow for more accurate predictions from the model. Interatomic potentials have been shown to be indispensable for studying many-particle systems and have allowed for the study of atomic processes that are not readily available in the lab \cite{PCC21}. This means we can conduct atomistic calculations of the force and quadratic coupling constants. In fact, these constants will be determined through the elements of the submatrices that make up the dynamical matrix given by
\begin{align}
\textbf{D}=
\begin{bmatrix}
\Phi_{ii}^{ab} & \Phi_{ij}^{ab} & \Phi_{ik}^{ab} \\
\Phi_{ji}^{ab} & \Phi_{jj}^{ab} & \Phi_{jk}^{ab} \\
\Phi_{ki}^{ab} & \Phi_{kj}^{ab} & \Phi_{kk}^{ab} 
\end{bmatrix} \label{DynamicalMatrix}
\end{align}
$i,j,k$ represent the x-, y-, and z-coordinates respectively. The constant $K$ will be determined by the diagonal submatrices ($\Phi_{ii}^{ab}, \Phi_{jj}^{ab}, \Phi_{kk}^{ab}$), while $G$ determined by the off-diagonal submatrices.

\begin{figure}[htbp]
\centering
\includegraphics[width=\columnwidth,scale=0.25]{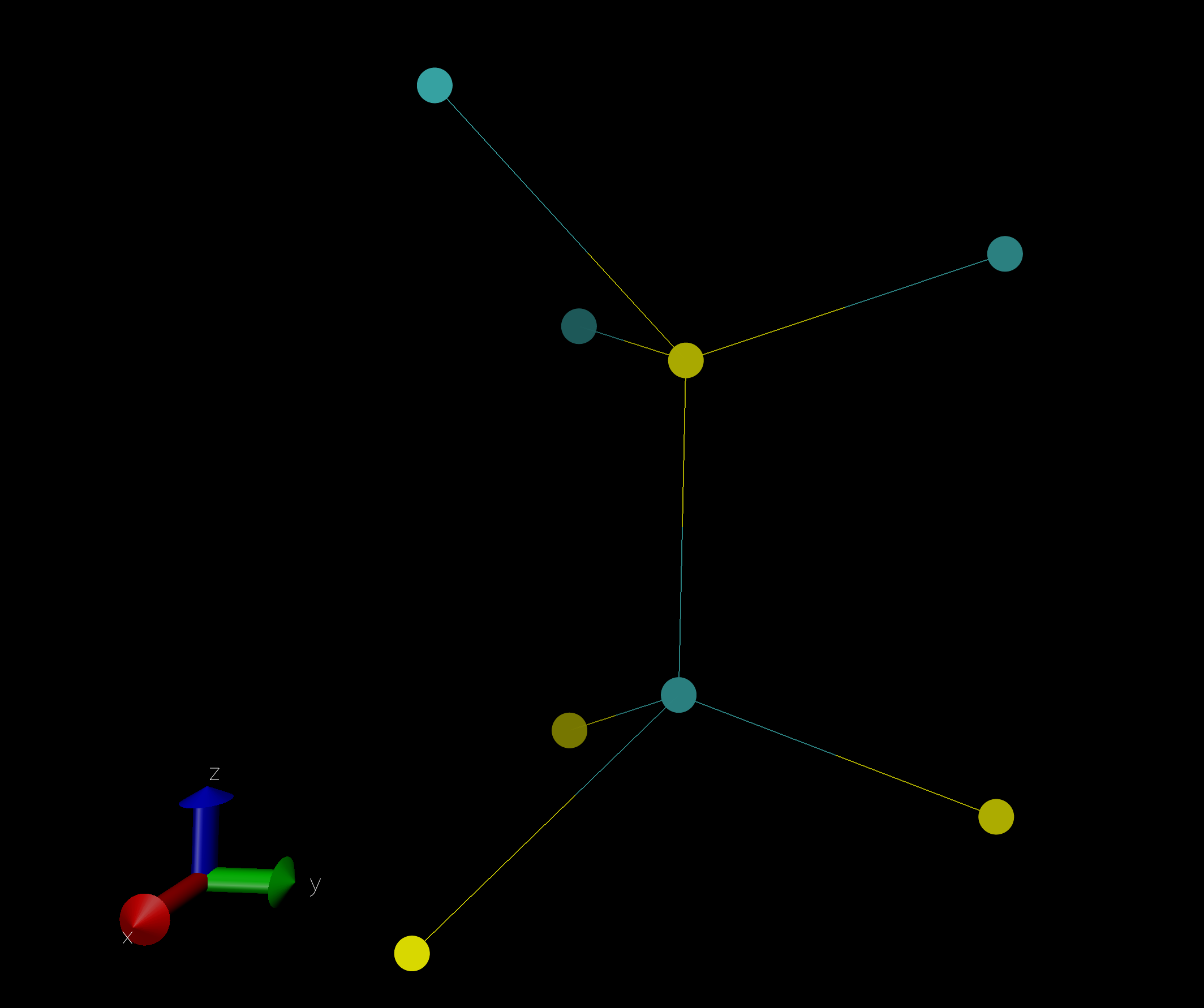}
\caption{The 8-atom unit cell structure section where a vacancy center can arise for the polytype $4H-$SiC used to calculate the baseline dynamical matrix. The green dots are carbon atoms and the yellow dots are silicon atoms. (This image was made with VMD software support. VMD is developed with NIH support by the Theoretical and Computational Biophysics group at the Beckman Institute, University of Illinois at Urbana-Champaign.)}
\label{atompositions}
\end{figure}

To conduct the atomistic calculations, we used Sandia National Laboratories open-source Large-scale Atomic/Molecular Massively Parallel Simulator (LAMMPS) \cite{TAB22}. LAMMPS lets us build a desired atomic structure environment, specify an interatomic potential, and perform a wide range of molecular dynamics calculations. To justify the choice of interatomic potential and the simulation environment is accurately modeling what has been seen experimentally, we test the section of the unit cell of $4H-$SiC where the vacancy center can arise. Since we are only concerned about the JT distortion within the vacancy center, we do not need to consider a supercell or try to generate a large atom collection of a possible concentration of many defect centers. We use techniques for studying isolated atom collections done previously using LAMMPS \cite{PCC21}. We test the interatomic potential without the vacancy to get accurate densities and bond strengths. Once our simulation can accurately calculate those values we can move to modeling $V_{Si}$ and $V^{\pm}_{C}$. The section of the unit cell atom structure (made up of 8 atoms) we are interested in for the simulation environment can be seen in Fig. (\ref{atompositions}).

For SiC (and in particular, $4H$-SiC), expansion of the environment-dependent interatomic potential (EDIP) \cite{BK96,BKJ97} for SiC has led to predictions agreeing with experiment \cite{LBP10,JMS12}. Lucas, \textit{et al} \cite{LBP10} described point defects of the disordered phase. Building upon this work, Jiang, \textit{et al} \cite{JMS12} used the corrected EDIP to accurately describe photoluminescence that was then seen experimentally. Due to the accuracy of EDIP, we use it here for the atomistic calculations of the vibronic nature of the $4H$-SiC polytype. 

\section{Results and Discussions}
\subsection{$F$, $K$, and $G$}
\begin{figure}[htbp]
\centering
\includegraphics[width=\columnwidth]{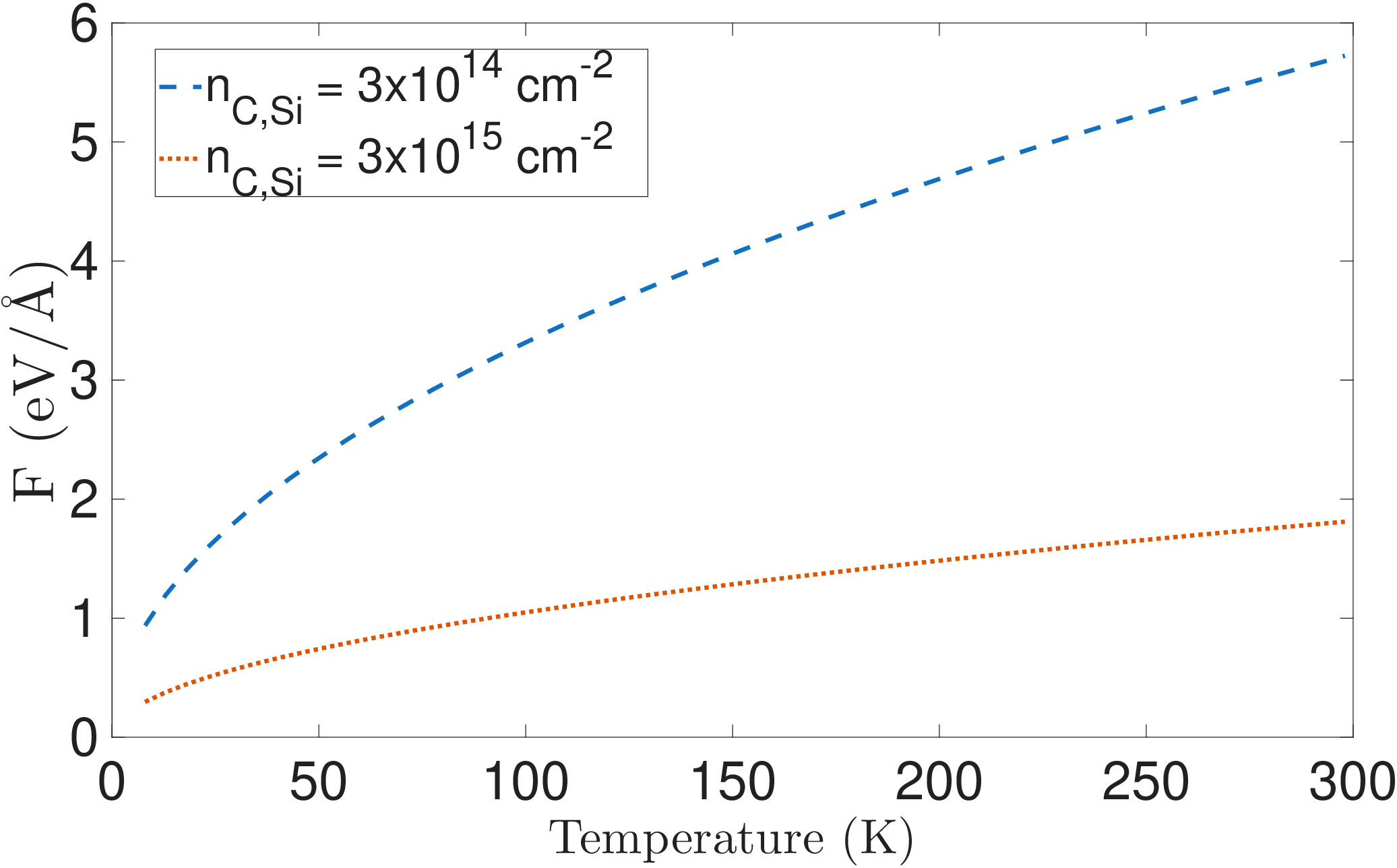}
\caption{Linear vibronic coupling, $F$, as a function of temperature. Blue dashed line is for $n_{C,Si} = 3 \times 10^{14} \mathrm{cm}^{-2}$ and red dotted is for $n_{C,Si} = 3 \times 10^{15} \mathrm{cm}^{-2}$. $F$ increases with increasing temperature, which is expected for semiconductors. At every temperature, a higher concentration of defects entails a decrease in the total number of coupled atoms, resulting in a lower $F$ value.}
\label{FTempDependence}
\end{figure}

With the model we have developed, we first want to calculate the coupling constants necessary to determine the APES to get $E_{JT}$. We first calculate the linear vibronic coupling, $F$, since we have an analytical method to determine its temperature dependence. From (\ref{Fanalytic}), we see that the main difference between $V_{Si}$ and $V^{\pm}_{C}$ comes from the concentration of the vacancies. Embley, \textit{et al} determined a $V_{Si}$ concentration of roughly $3 \times 10^{14} \mathrm{cm}^{-2}$ to $3 \times 10^{15} \mathrm{cm}^{-2}$ depending on how irradiated their sample was \cite{ECM17}. In Fig. (\ref{FTempDependence}), we assume these concentrations and see a steady increase in $F$ as temperature increases. This is expected for semiconductors like $4H-$SiC. There is also a decrease in coupling with increasing concentration of vacancies. This comes from Eqn. (\ref{Fanalytic}), but, also makes intuitive sense: more defects means fewer total atoms to couple, lowering the total vibronic coupling. Our low temperature linear couplings fall within the same order of magnitude of the values seen from DFT models for the divacancies in $4H-$SiC \cite{ZSV24} giving confidence in our model.

\begin{figure*}[htbp]
    \begin{subfigure}{0.98\columnwidth}
        \includegraphics[width=\columnwidth,scale=0.25]{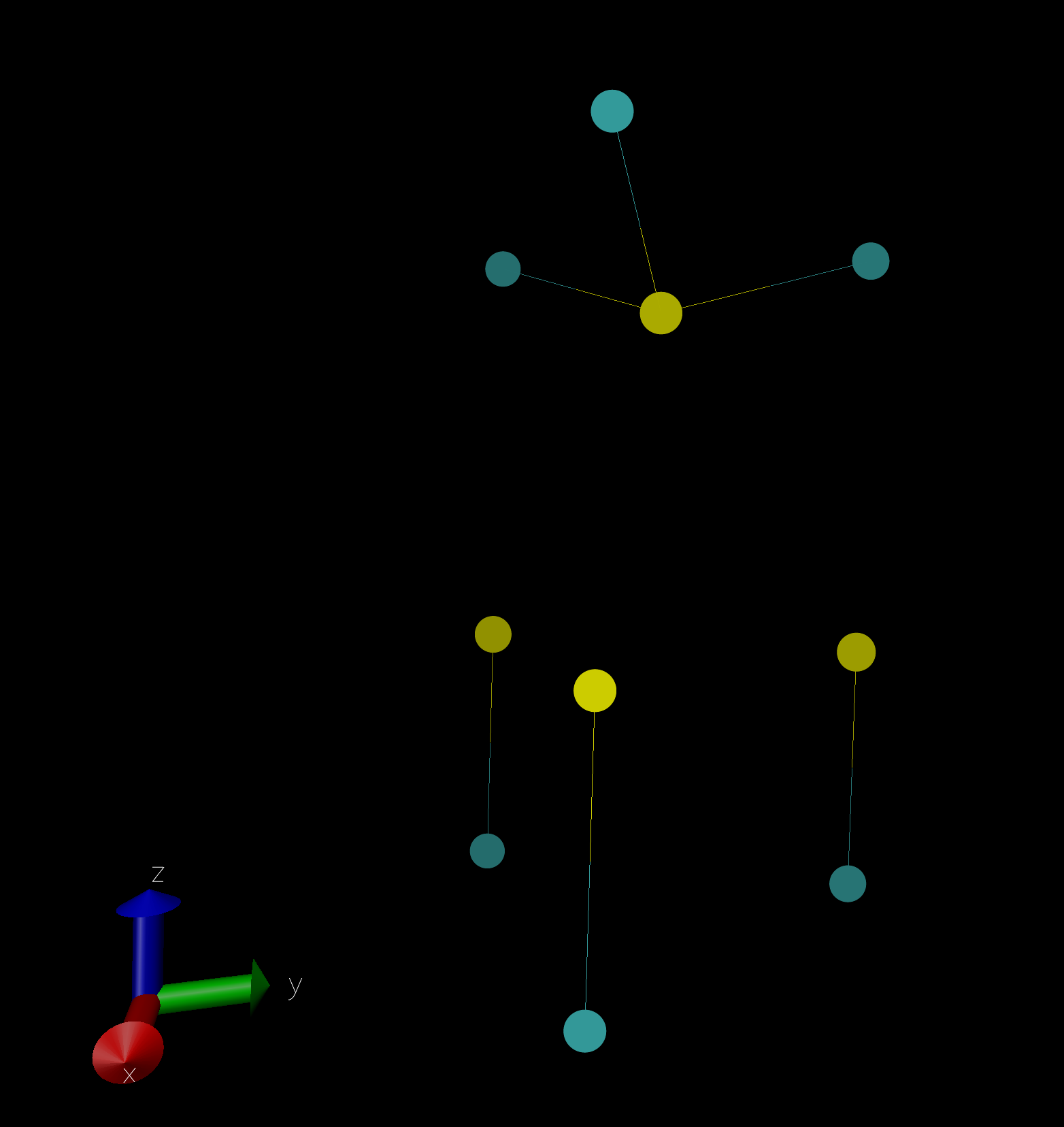}
        \caption{}
        \label{CVacancy}
    \end{subfigure}
    \begin{subfigure}{\columnwidth}
        \includegraphics[width=\columnwidth,scale=0.25]{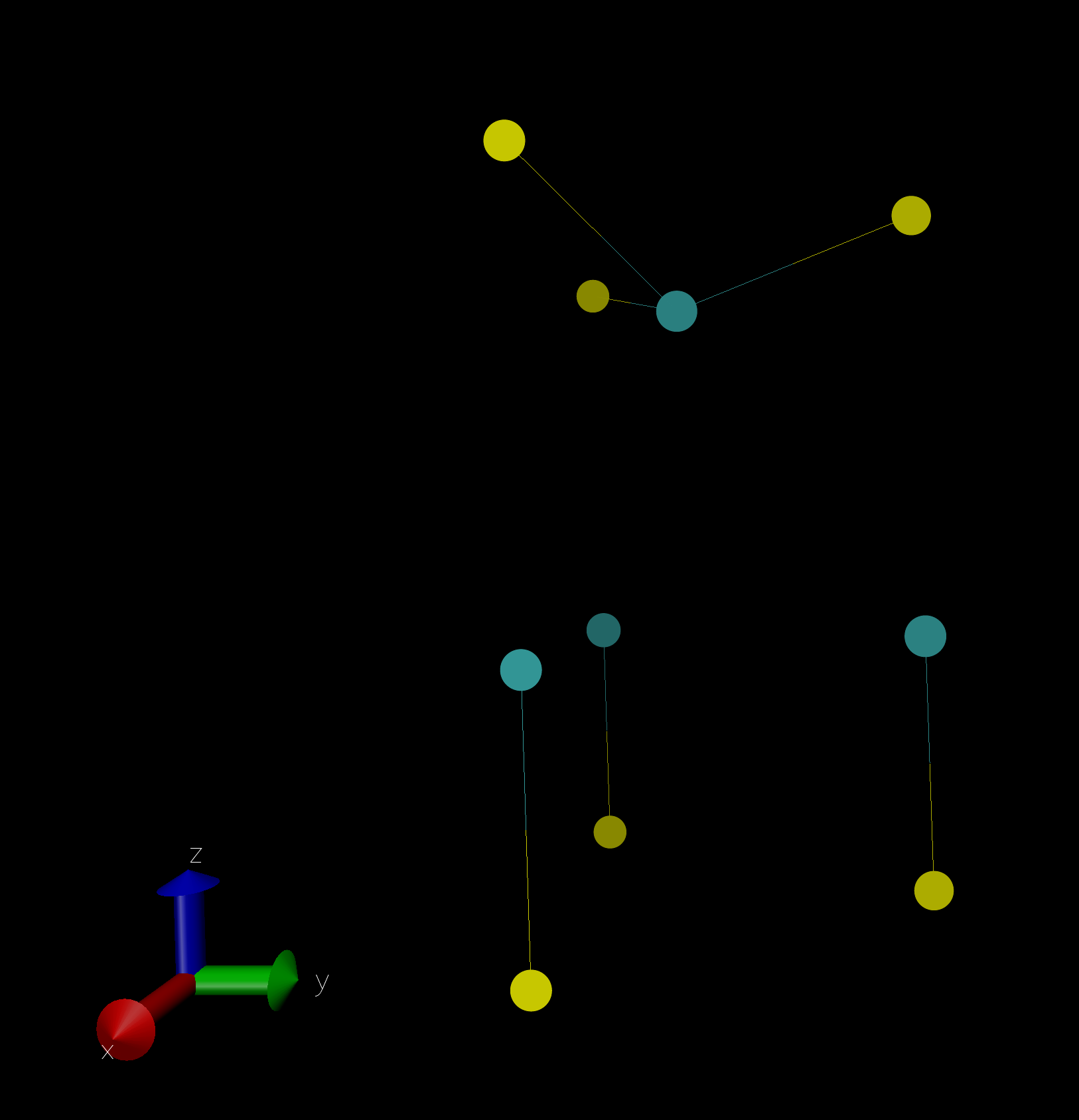}
        \caption{}
        \label{SiVacancy}
    \end{subfigure}
     \caption{The simulation environment made up of 10 atoms for the a.) silicon and b.) carbon vacancy ($V_{Si},V_{C}$) of the polytype 4H-SiC to calculate the dynamical matrix. The green dots are carbon atoms and the yellow dots are silicon atoms. (This image was made with VMD software support. VMD is developed with NIH support by the Theoretical and Computational Biophysics group at the Beckman Institute, University of Illinois at Urbana-Champaign.)}
\end{figure*}

The $4$H-SiC polytype can have two possible vacancies: the positively and negatively charged carbon ($V^{\pm}_{C}$) or silicon ($V_{Si}$) vacancy. We started our atomistic calculations of these vacancy centers by verifying that our simulation environment can reproduce unit cell density and bond strengths. This gives an initial calibration before moving to atomistic calculations needed to determine $K$ and $G$. The simulation environment for the vacancy centers is seen in Figs. (\ref{CVacancy}) and (\ref{SiVacancy}).

After calibrating the simulation environment, we are able to calculate the dynamical matrix and extract the force constant ($K$) and quadratic coupling constant ($G$) for each of the vacancy centers. For the $V_{Si}$ center we get a $K$ value of 0.9614 eV/\r{A}$^{2}$ and a $G$ value of 0.0425 eV/\r{A}$^{2}$. Alternatively, for the $V^{\pm}_{C}$ center we obtained a $K$ of 0.9837 eV/\r{A}$^{2}$ and a $G$ value of 0.0994 eV/\r{A}$^{2}$. These values are given in Table (\ref{couplingconstantvalues}). We can also see that when compared to the divacancy values calculated with DFT methods by Zalandauskas \textit{et al} \cite{ZSV24}, our values are within the same order of magnitude giving confidence in their accuracy.
\begin{table}[htbp]
\begin{center}%
\begin{tabular}
[c]{|c|c|c|}
\hline
\multicolumn{3}{c}{\textbf{$K$ and $G$ Constants} (\textrm{eV}/\r{A}$^{2}$)} \\
\hline
\textbf{Vacancy} & \textbf{$K$} (eV/\r{A}$^{2}$) & \textbf{$G$} (eV/\r{A}$^{2}$)\\\hline
$V_{Si}$ & 0.9614 & 0.0425 \\\hline
$V^{\pm}_{C}$ & 0.9837 & 0.0994 \\\hline
\end{tabular}
\end{center}
\caption{The force constant ($K$) and quadratic coupling constant ($G$) as determined through atomistic calculations using environment-dependent interatomic potential (EDIP) semi-empirical potential for the silicon and carbon vacancies ($V_{Si}, V^{\pm}_{C}$).}
\label{couplingconstantvalues}
\end{table}

We note here that a comprehensive temperature-dependent calculation is not done for the constants $K$ and $G$ at this time. As these constants are still needed to calculate the APES, we will consider them constant similar to that done previously in DFT-based calculations of the APES \cite{ZSV24}. We can still calculate the temperature dependence of $E_{JT}$ due to the fact that we can calculate the temperature dependence of $F$ which is also crucial to calculating the APES (see Eqn. (\ref{APES})). At this time, an in-depth study of the temperature dependence of the coupling constants $K$ and $G$ is out of the scope of this current work as it would require extensive validation by studying a wide range of semi-empirical potentials or temperature-dependent DFT as no temperature dependent studies of these vibronic coupling have been done thus far based on current literature. Using our temperature dependent model for $F$ and our atomistically calculated values for $K$ and $G$, we can estimate the $E_{JT}$ for the $V_{Si}$ and $V^{\pm}_{C}$ single vacancy centers and compare to what has been calculated for the divacancies of $4H-$SiC.

\subsection{APES and $E_{JT}$ calculation}
\begin{figure*}[htbp]
    \begin{subfigure}{\columnwidth}
        \includegraphics[width=\columnwidth]{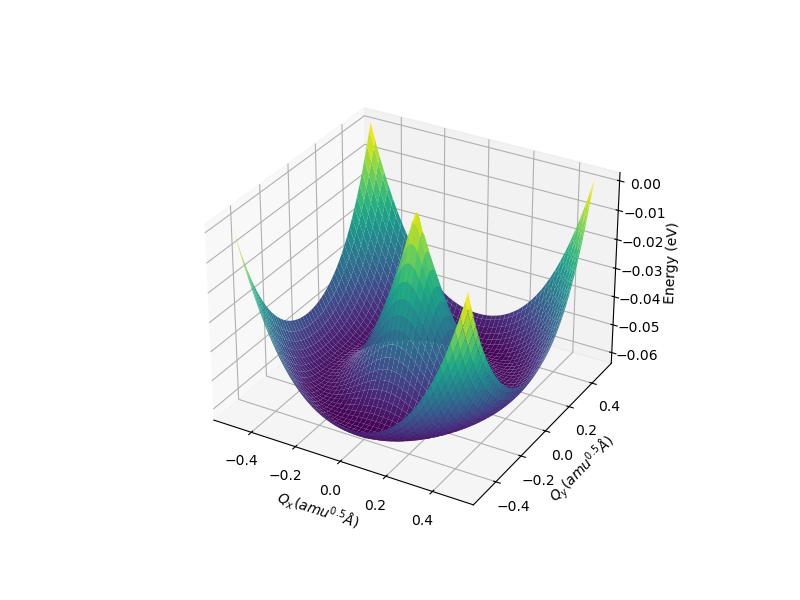}
        \caption{}
        \label{APES_SiVacancy}
    \end{subfigure}
    \begin{subfigure}{\columnwidth}
        \includegraphics[width=\columnwidth]{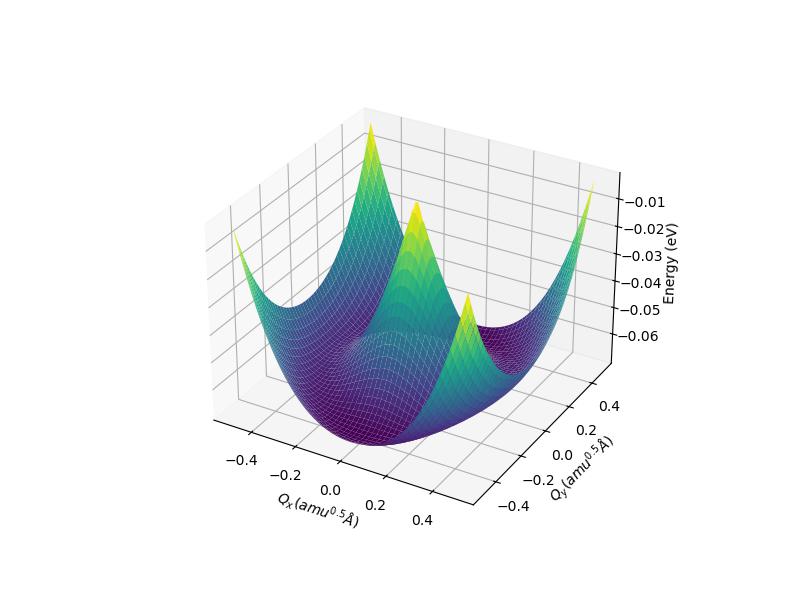}
        \caption{}
        \label{APES_CVacancy}
    \end{subfigure}
    \caption{The near zero Kelvin adiabatic potential energy surface for the a.) silicon vacancy ($V_{Si}$) and b.) carbon vacancy ($V^{\pm}_{C}$). The mexican hat shape can be seen which allows the determination of the Jahn-Teller stabilization energy ($E_{JT}$) and the energy barrier $\delta_{JT}$.}
\end{figure*}

With $K$ and $G$ calculated atomistically and $F$ calculated analytically, we can generate the APES at the near zero-temperature limit from Eqn. (\ref{APES}). The APES for $V_{Si}$ and $V^{\pm}_{C}$ at the near zero-kelvin limit can be seen in Figs. (\ref{APES_SiVacancy}) and (\ref{APES_CVacancy}) respectively. We consider the same value for $F$ for both systems for brevity. From the APES we can extract $E_{JT}$ by taking the difference between the energy of the peak at the origin ($Q=0$) and the valley (the darkest blue area) as shown within Fig. (\ref{APES_SiVacancy}). We also are able to determine the energy barrier, $\delta_{JT}$, by taking the difference of the energy at the base of the cone to the valley.

In Figs. (\ref{APES_SiVacancy}) and (\ref{APES_SiVacancy}), we have graphed the APES for $V_{Si}$ and $V^{\pm}_{C}$. We see that $V^{\pm}_{C}$ gives a larger $E_{JT}$ than $V_{Si}$ of $E_{JT}^{C}=71.1$ meV compared to $E_{JT}^{Si}=62.7$ meV. The higher $E_{JT}$ for $V^{\pm}_{C}$ gives an overall larger decrease in stabilization energy from the spontaneous symmetry lowering. The larger energy decrease for the $V^{\pm}_{C}$ is likely due to the larger couplings of $G$ and $K$. We also get a value of $\delta_{JT}^{Si}=10.2$ meV and $\delta_{JT}^{C}=23.5$ meV for the $V_{Si}$ and $V^{\pm}_{C}$ defects respectively. Compared to $E_{JT}$ and $\delta_{JT}$ for the divacancies in $4H-$SiC, our results are within the same order of magnitude. Since the divacancy is the removal of both a silicon and a carbon atom, we look at the average $E_{JT}$ from our $V_{Si}$ and $V^{\pm}_{C}$ single vacancies. We see that their average comes out to be $66.9$ meV. If we do the same for $\delta_{JT}$, we get an average contribution from each vacancy of $16.85$ meV. Our calculated $E_{JT}$ is consistent with the DFT-based calculations of Zalandauskas \textit{et al}, who found $69$ meV and $64$ meV for the divacancy and our $\delta_{JT}$ is close to their $23$ meV and $19$ meV. Overall, we have confidence in the JT stabilization energies we calculated using analytical and atomistic methods for $F$, $K$, and $G$. With confidence in our calculations of $E_{JT}$, we use Eqn. (\ref{Fanalytic}) with Eqn. (\ref{APES}) to calculate $E_{JT}$ at different temperatures.

\subsection{$T_{2}^{JT}$ temperature dependence}
\begin{figure}[htbp]
\centering
\includegraphics[width=\columnwidth]{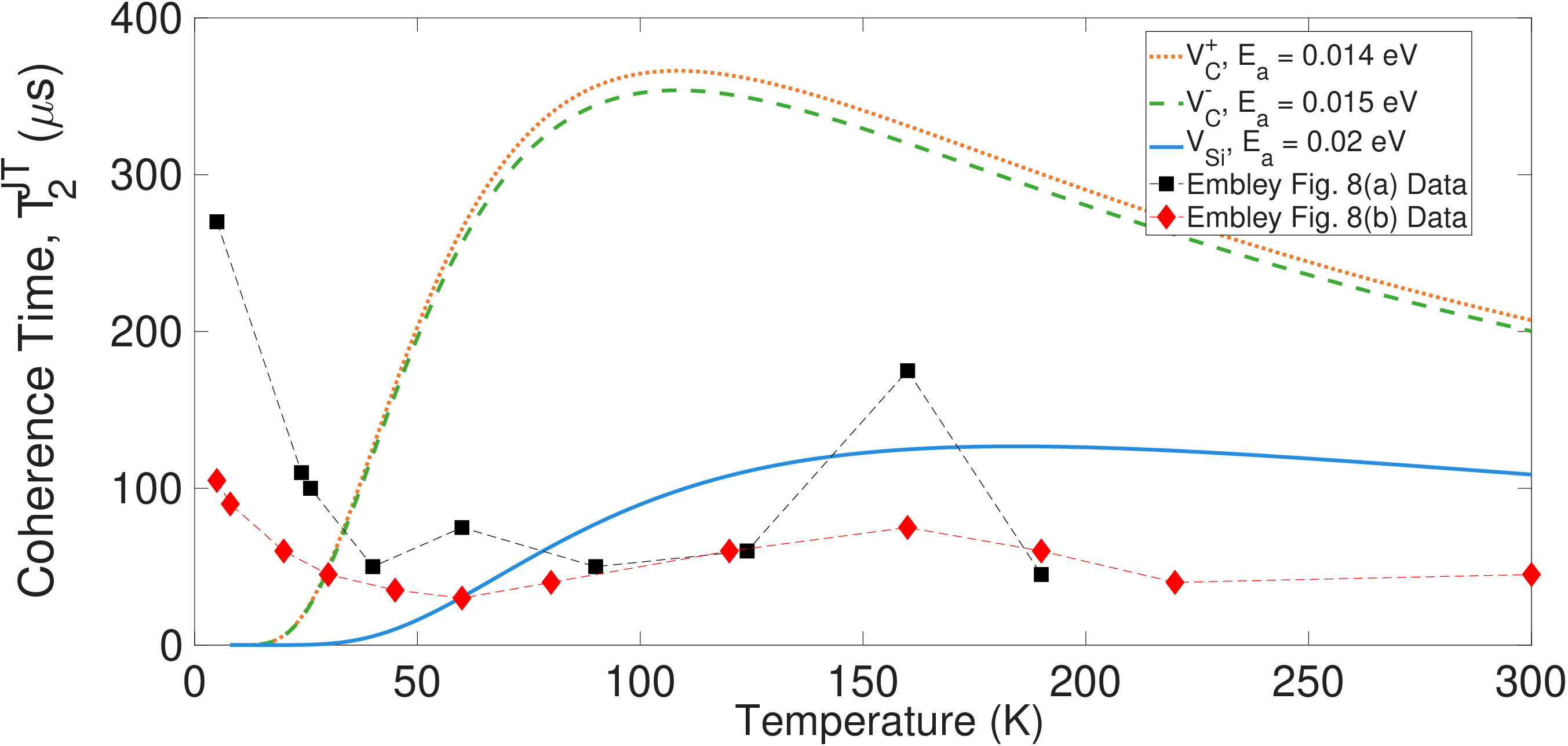}
\caption{The coherence time due to the motional Jahn-Teller distortion versus temperature for the positively ($V_{C}^{+}$, orange dotted), negatively charged carbon ($V_{C}^{-}$, green dashed) vacancy, silicon vacancy ($V_{Si}$, solid blue), and Embley \textit{et al} Fig. 8(a) (black squares) and Fig 8(b) (red diamonds) experimental data \cite{ECM17}. Here the activation energy of for $V_{Si}$ is $E_{a}=20$ meV, $V_{C}^{+}$ is $14$ meV, and $V_{C}^{-}$ $15$ meV). At low temperatures, our model predictions for the $V_{Si}$ shows a rise earlier than the experimental data of Embley \textit{et al}. Whereas the peak lines up well. The reasoning is due to phonon-assisted relaxation explained in Sec. (\ref{PhononIncorporated})}
\label{T2_JT_TempDependence}
\end{figure}

With $E_{JT}$ determined, we can plug it back into Eqn. (\ref{LarmMagField}) and graph the coherence time temperature dependence as described by equation (\ref{T2Complete}). In Fig. (\ref{T2_JT_TempDependence}), we graph the coherence time for the $V_{Si}$ and $V^{\pm}_{C}$ single vacancies due to the motional JT distortion only. We do this to isolate the effect on coherence due to the JT distortion. Remembering that the motional narrowing effect due to the polaron hopping is thermally activated with a specific activation energy, we adopt Umeda \textit{ et al}'s activation energy of $14$ meV for the vacancies $V_{C}^{\pm}$ \cite{UIM04,UII05} and Watkins \textit{et al}'s $20$ meV for $V_{Si}$ \cite{WC64,WC65,GDW68}. 

Fig. (\ref{T2_JT_TempDependence}) shows that when the temperature reaches $20$K for the $V_{C}^{\pm}$ single vacancies, the coherence time begins to increase. This increase of the coherence time is the motional narrowing effect taking place (i.e., "turning on"). As temperatures increase to $120$K, the coherence peaks and is followed by a gradual decrease. At this point, Larmor fluctuations start overtaking the motional narrowing effect (i.e., motional narrowing "turning off"). For $V_{Si}$, we see that around $40$K the coherence time starts to increase, then it peaks around $150$K. This is in good agreement with what has been observed experimentally \cite{UIM04,UII05,ECM17}. We note that Embley \textit{et al} \cite{ECM17} noticed a rise in coherence times closer to $60$K and a peak closer to $160$K for $V_{Si}$. In Sec. \ref{PhononIncorporated}, we incorporate the phonon-assisted spin relaxation to reproduce their findings more accurately.

The increase in coherence time comes from motional narrowing due to the thermally activated polaron hopping. This hopping (that is, electronic bond switching) occurs between Si$_{1}$-Si$_{2}$, Si$_{1}$-Si$_{3}$, and Si$_{1}$-Si$_{4}$ for the $V_{C}^{\pm}$ vacancies and C$_{1}$-C$_{2}$, C$_{1}$-C$_{3}$, and C$_{1}$-C$_{4}$ for the $V_{Si}$ vacancy. Because the defect is paramagnetic, the nuclear spins are randomly oriented and naturally with increasing temperature will have the overall magnetization decrease which is Curie's law for paramagnetics. The increase in polaron hopping rate then leads to a decrease in overall magnetization felt by the vacancy center electron spin. These different physical mechanisms become the catalysts for motional narrowing to take place and the coherence time to increase. 

\begin{figure*}[htbp]
    \begin{subfigure}{\columnwidth}
    \includegraphics[width=\columnwidth]{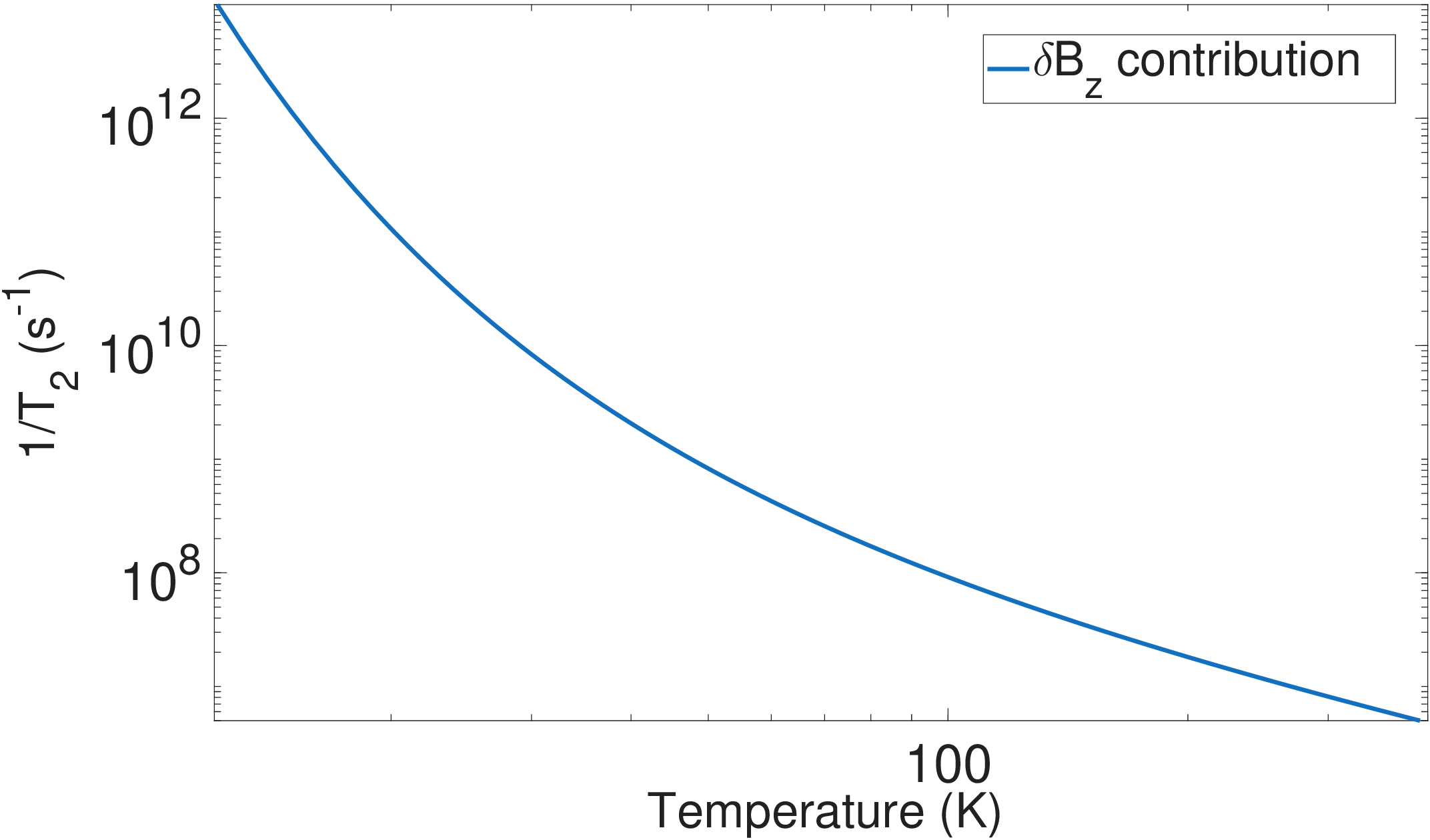}
    \caption{}
    \label{T2_TempDependence}
    \end{subfigure}
    \begin{subfigure}{\columnwidth}
    \includegraphics[width=\columnwidth]{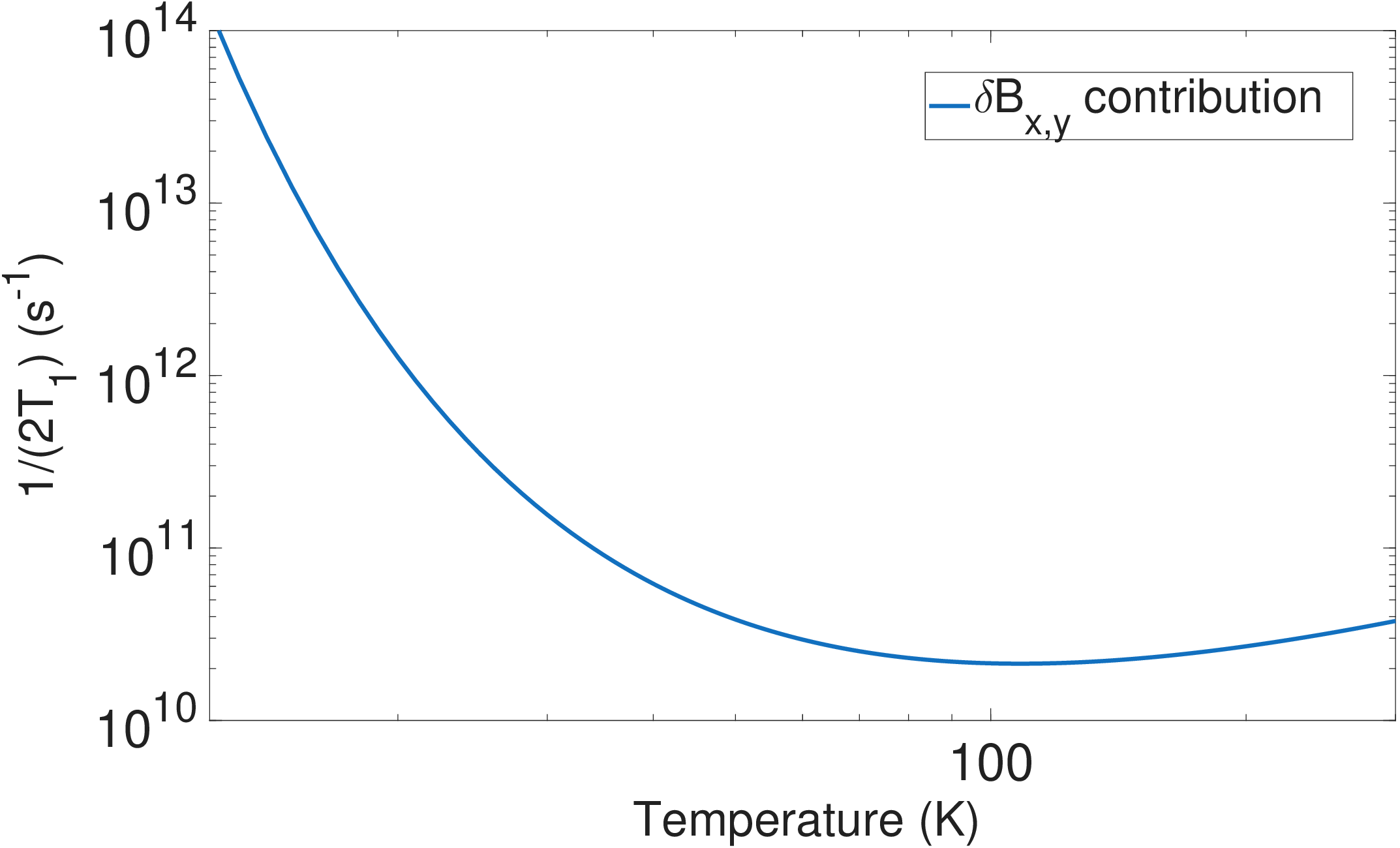}
    \caption{}
    \label{T1_TempDependence}
    \end{subfigure}
    \caption{a.) The effect of the $\delta B_{z}$ magnetic field fluctuations due to randomly oriented nuclear spins. As temperature increases, the fluctuations of the magnetic field decrease, which follows Curie's law for paramagnetics as expected. b.) The effect of the $\delta B_{x,y}$ magnetic field fluctuations due to Larmor precession and polaron hopping. As the temperature increases, the fluctuations of the magnetic field decrease until a minimum around $100-120$K due to the polaron hopping, resulting in a minimum net magnetization. After the minimum, the fluctuations increase as the polaron hopping slows down and Larmor precession dominates.}
\end{figure*}

The thermally driven activation energy of the polaron hopping dictates when the motional narrowing begins (or "turns on") and peaks (or "turns off"). The peaking of the coherence comes from the decay of the hopping rate due to an increasing JT energy barrier. The decoherence is then worsened as Larmor precession fluctuations begin to dominate. Figs. (\ref{T1_TempDependence}) and (\ref{T2_TempDependence}) show the switching from an only reducing magnetic field fluctuation to a fight between a decreasing and increasing magnetic field fluctuation. The initial decrease in magnetic field fluctuations allows the motional narrowing effect to take place (narrowing of the spectral linewidth; increase in coherence). Once temperatures get high enough, the Larmor precession begins to dominate, increasing the magnetic field fluctuations, which then causes linewidth broadening (i.e., decoherence). Watkins, \textit{et al.} attributed the bond switching distortion to the redistribution of the electron cloud \cite{WC64,WC65,GDW68}. The redistribution of the electron cloud can also be described by polaron hopping. Polarons naturally localize the electron, changing its overall wave function, which affects the overall electron cloud \cite{BCS24,FRS21}.

In Fig. (\ref{T2_TempDependence}), the $\delta B_{z}^{2}$ term of Eqn. (\ref{T2Complete}) decreases with temperature as expected for a paramagnetic material. On the other hand, in Fig. (\ref{T1_TempDependence}), the $\delta B_{x}^{2} + \delta B_{y}^{2}$ term initially decreases until it reaches a minimum, then begins to increase. At this point, the Larmor precession will start to dominate the magnetic field fluctuations. As the polaron hopping slows due to an increasing energy barrier established by the decreasing JT distortion at higher temperatures, the magnetic field fluctuations start to increase leading to decoherence and a decreasing coherence time.

\subsection{Incorporating phonon-assisted relaxation, $T_{1}^{phonon}$} \label{PhononIncorporated}
At low temperatures, the polaron thermal energy is not sufficient for hopping to begin. As a result, at low temperatures, polaron hopping does not contribute to the overall coherence time of Eqn. (\ref{T2Complete}). At these low temperatures ($<20$K), though, phonon-assisted spin-lattice relaxation $T_{1}^{phonon}$ is significant. Simin, \textit{et al} \cite{SKS17} studied phonon-assisted spin relaxation due to the thermal bath for $V_{Si}$, and found a power law dependence \cite{SKS17},
\begin{align}
\frac{1}{T_{1}^{phonon}(T)} = A_{0} + A_{1}T + A_{5}T^{5} + \frac{R}{e^{\Delta/k_{B}T}-1}. \label{T1phonon}
\end{align} 
Here, $A_{0,1,5}$ and $R$ are constants and $\Delta$ is the energy of the local phonon mode at the $V_{Si}$ defect. We can then incorporate Eqn. (\ref{T1phonon}) into a $T_{2}^{*}$ calculation.

\begin{figure}[htbp]
\centering
\includegraphics[width=\columnwidth]{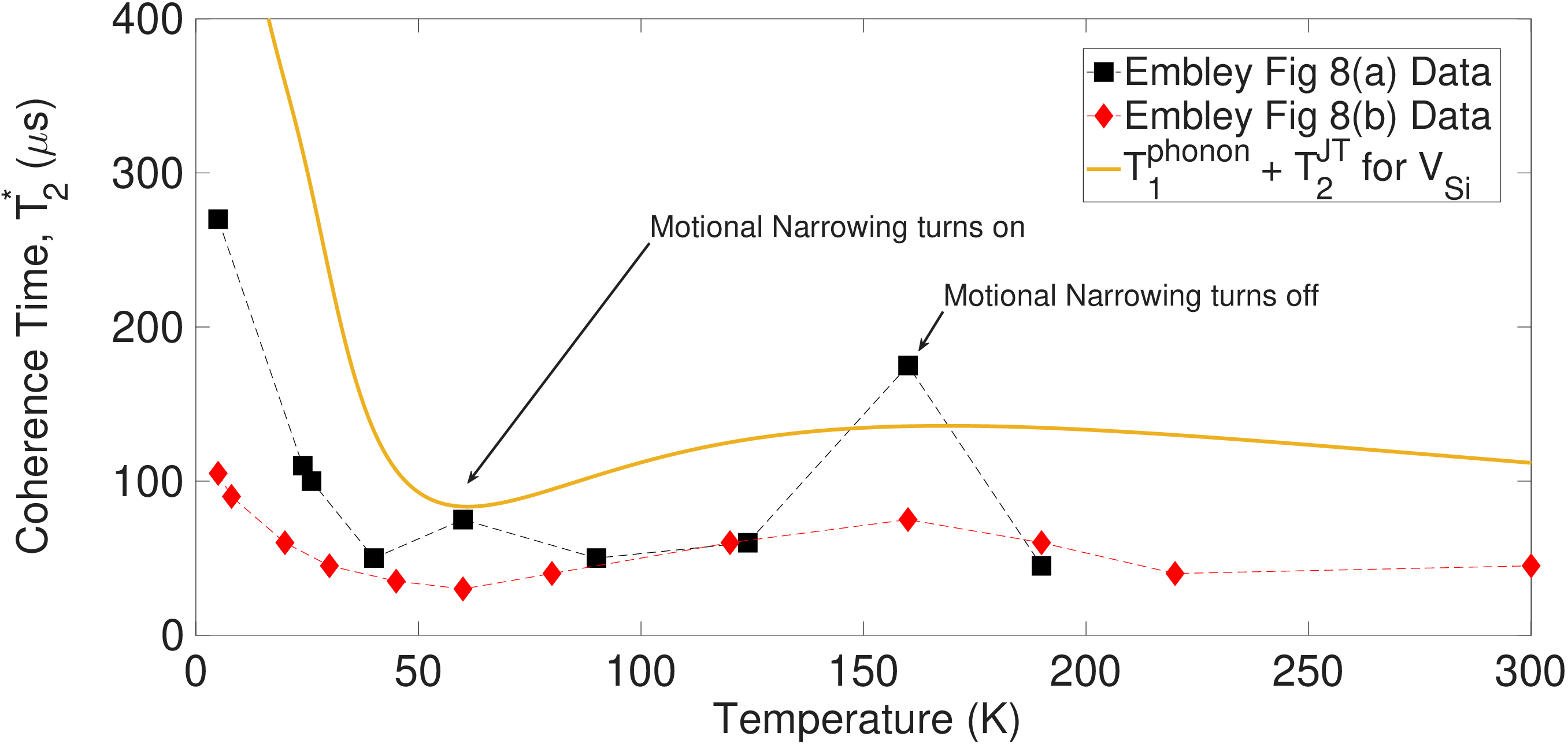}
\caption{The coherence time, $T_{2}^{*}$, due to both the phonon-assisted spin relaxation and the motional Jahn-Teller distortion versus temperature for the silicon vacancy ($V_{Si}$) (Orange Solid). The data from Fig. 8(a) (black squares) and 8(b) (red diamonds) of Embley \textit{et al} are plotted here as well \cite{ECM17}. Our model calculation shows a minimum around $60$K where motional narrowing turns on. As temperatures reach the order of $160-170$K, the polaron hopping starts slowing down leading to the motional narrowing turning off at the peak. This calculation is in good agreement with what was seen experimentally by Embley \textit{et al} \cite{ECM17}.}
\label{T2WithPhonon}
\end{figure}

Fig. (\ref{T2WithPhonon}) plots the coherence time $T_{2}^{*}$ as a function of temperature for $V_{Si}$. Both $T_{2}^{JT}$ and $T_{1}^{phonon}$ from Eqn. (\ref{T2WithPhonon}) contribute to $T_{2}^{*}$. Incorporating the thermal bath at low temperatures, we are able to reproduce with great agreement the coherence time temperature dependence observed experimentally by Embley \textit{et al} \cite{ECM17} (black squares and red diamonds in Fig. (\ref{T2WithPhonon})). We see that for temperatures $<60$K, phonon-assisted spin relaxation dominates coherence dephasing. As temperatures reach $60$K, the thermal activation of polaron hopping turns on the motional narrowing effect. This causes the coherence time to be dominated by the decreasing total magnetization seen by the polaron. At these lower temperatures, the system has both terms in Eqn. (\ref{T2Complete}) giving a decreasing magnetic field fluctuation. As a result, coherence time increases and the motional narrowing effect has started. Once temperatures approach $160-170$K, the energy barrier due to the decreasing JT distortion becomes high enough to slow the polaron hopping. With the polaron hopping slowing down, the magnetic field fluctuations become dominated by Larmor precession, decreasing the overall coherence time and effectively "turning off" the motional narrowing effect. 

We can explain the difference between Figs. (\ref{T2_JT_TempDependence}) and (\ref{T2WithPhonon}) by the addition of the phonon-assisted spin relaxation. The thermal activation of the motional narrowing shifts from $40$K to $60$K due to the electron spin interacting with the phonon cloud. Once the polaron hopping gets fast enough, spin relaxation from the phonon cloud becomes negligible, allowing the motional narrowing effect to "turn on" completely. We currently do not incorporate the other noise sources that would be prominent, especially as temperatures approach room temperature. Even so, as can be seen in Figs. (\ref{T2_JT_TempDependence}) and (\ref{T2WithPhonon}), we are still able to replicate with great accuracy what is seen experimentally: this is a significant validation of our model.

Since defect centers such as the $V^{\pm}_{C}$ and $V_{Si}$ and divacancies of $4$H-SiC and $6$H-SiC and that of the NV center are prime for the generation of a polaron, there is a high probability that motional narrowing will occur \cite{LP08}. In fact, a similar quasiparticle known as the polariton, which is generated from a coupling between a photon and matter excitations (phonon or exciton), can also explain the observation of motional narrowing in semiconductor quantum systems \cite{WET21}. In systems where a polaron, polariton, or other similar quasiparticle is generated, motional narrowing is likely to occur. The motional narrowing onset temperature depends on the thermal energy barrier, that is, the thermal activation energy $E_{a}$. So far, it seems that no systematic measurement of $T_{2}$ at different temperatures, similar to the $8$-$300$~K range studied by Embley, \textit{et al.}, has been made for NV centers. Also, only recently have researchers started looking into polarons within these different defect centers \cite{UTM20,IGF24}.

Polaron formation and its link to Jahn-Teller distortion has been widely studied \cite{TKB04,SBT03,YT00,PEK00,LCK03}. Polarons are also present in a wide range of applications where motional narrowing of different types has been observed \cite{SCL19,UNK82,MHT08,ZBL24,NSP24}. As our calculations show, the motional narrowing that takes place can be described by the generation of a polaron in $V_{C}^{\pm}$ and $V_{Si}$ single vacancy defects of $4H$-SiC. A similar motional narrowing has also been seen in $6H$-SiC polytype divacancies through EPR \cite{BMO01} and direct coherence measurements \cite{FKI15}. This motional narrowing can also be described through the generation of a polaron---specifically, a bipolaron. It has been shown that bipolarons also lead to motional narrowing in spin systems \cite{HH94,LSM92,HPS90}.

\section{Conclusions}
Because orbital degeneracy leaves the vacancy defect vulnerable to spontaneous symmetry lowering, motional JT distortion occurs, lifting the degeneracy. The vibronic coupling then generates a polaron quasiparticle. We follow an analytical approach and build an atomistic model of the vibronic coupling parameters to calculate the JT stabilization energy $E_{JT}$ and energy barrier $\delta_{JT}$ for the single vacancy centers $V_{Si}$  and $V^{\pm}_{C}$. To validate our results for $E_{JT}$ and $\delta_{JT}$ at near-zero Kelvin, we compare to previous work using DFT simulations of the $4$H-SiC divacancies \cite{ZSV24}. We see our values for $E_{JT}$ and $\delta_{JT}$ from out atomistic model are of the same order of magnitude as the DFT data. Also, if we assume our values for $E_{JT}$ and $\delta_{JT}$ contribute equally in a divacancy, our calculations are in good agreement with the results of DFT-based calculations \cite{ZSV24}.

For our $T^{JT}_{2}$ calculations as seen in Fig. \ref{T2_JT_TempDependence}, we see for $V^{\pm}_{C}$ as the temperature reaches $20$K, the thermal energy is large enough to activate polaron hopping and motional narrowing to "turn-on" and then "turn-off" as the energy barrier gets to high at $120$K. This is in great agreement with EPR observations \cite{UIM04,UII05}. As for $V_{Si}$, we see the "turning on" of motional narrowing occur around $40$K and "turn-off" at $150$K which is in great aggreement with the systematic temperature experiments done by Embley \textit{et al} \cite{ECM17}. When incorporating the phonon-assisted relaxation and calculating $T^{*}_{2}$ for $V_{Si}$, we see motional narrowing "turn on" closer to $60$K and "turn-off" around $160$K. Comparing our calculation again to the experimental data of Embley \textit{et al} \cite{ECM17} as seen in Fig. \ref{T2WithPhonon}, there is even better agreement giving great validation for our model.  

In this work, we have theoretically investigated the coherence time anomaly seen in the $V^{\pm}_{C}$ and $V_{Si}$ single vacancies of $4H$-SiC. We have considered the motional Jahn-Teller (JT) distortion that occurs in the vacancy defect due to the orbital degeneracy making the vacancy center vulnerable to spontaneous symmetry lowering \cite{UTM20,IDS17,AG12,IBB06}. Our calculations reproduce the sudden and unexpected increase in coherence time with increasing temperature observed experimentally for the $V^{\pm}_{C}$ and $V_{Si}$ single vacancies \cite{ECM17,UIM04,UII05}. We have determined the mechanisms, not fully understood until now, that cause the increase in coherence time. Lastly, we have linked the coherence time increase with temperature to a universal effect.

Future work into a fully temperature dependent atomistic model of the $V^{\pm}_{C}$ and $V_{Si}$ single vacancies and the divacancies in $4$H-SiC and $6$H-SiC temperature dependent vibronic couplings will strengthen our model quantitatively. Recent theoretical studies using the cluster correlation expansion technique have looked into promising spin defect center hosts, which could have coherence times up to $47$~ms at room temperature \cite{KHS22}. Along with building a more complete atomistic model of the vibronic couplings, experiments systematically measuring the temperature dependence of $T_{2}$ could give observational evidence that the motional narrowing effect due to polaron quasiparticle generation is a universal phenomenon for these types of defect center spin systems.

\section{Acknowledgements}
This work is supported by the Advanced Quantum Sensing Center DoD Contract No. W911NF2020276. Sandia National Laboratories is a multimission laboratory managed and operated by National Technology and 
Engineering Solutions of Sandia, LLC, a wholly owned subsidiary of Honeywell International, Inc., for the DOE’s National Nuclear Security Administration under Contract No. DE-NA0003525. This paper describes objective technical results and analysis. Any subjective views or opinions that might be expressed in the paper do not necessarily represent the views of the U.S. Department of Energy or the United States Government. The authors thank Jonathan Tannenhauser for the valuable input.

\appendix
\section{Temperature Dependent Magnetic Fields Derivation\label{MagFieldTempDepDeriv}}
\subsection{General paramagnetism \label{Appdx:paramagnetism}}

In this work we are considering the vacancy centers ($V_{Si}$,$V_{C}$) in $4H$-SiC. Instead of starting there, let us first consider an electron from a paramagnetic material at some temperature, $T$. The electron will see an external magnetic field $\textbf{B}=B_{z}\textbf{e}_{z}$ from the nuclei it is bound to. Here we will have the nuclear field point in the $z$-direction. The magnetic energy can then be written as,
\begin{equation}
\epsilon = -\mathbf{\mu}\cdot\textbf{B},
\end{equation}
where $\mu$ is the magnetic moment. The magnetic moment of an atom is directly proportional to the total electronic angular momentum ($\textbf{J}$) given by $\hbar\textbf{J}$, where $\hbar$ is Planck's constant. This proportionality is fully given by
\begin{equation}
\mathbf{\mu} = g\mu_{B}\textbf{J},
\end{equation}
where $\mu_{B}=e\hbar/2m_{e}$ is the Bohr magneton and $g$ is the electron g-factor. Here, $e$ is the electron charge and $m_{e}$ is the mass of the electron. We can now write the magnetic energy as
\begin{equation}
\epsilon = -g\mu_{B}\textbf{J}\cdot\textbf{B}.
\end{equation}
According to quantum mechanics, $\textbf{J}$ can take only positive integer or half-integer values. Depending on $\textbf{J}$, the atom can be in specific magnetic energy levels. For instance, a spin-$1/2$ particle can be in a magnetic energy level $m=\pm1/2$.

The probability of the atom being in a certain magnetic level can be written as
\begin{equation}
P_{m}=\mathrm{exp}\left(\frac{g\mu_{B}B_{z}m}{k_{B}T}\right),
\end{equation}
where $k_{B}$ is Boltzmann's constant. We are considering the nuclear magnetic field along the z-axis. From this, we can write the magnetic moment along the z-axis as
\begin{equation}
\mu_{z} = g \mu_{B}m.
\end{equation}
We can then write the mean magnetic moment along the z-axis based on the probability of the atom being in one of the magnetic sublevels,
\begin{equation}
\overline{\mu_{z}} = \frac{\sum_{m=-J,J}\mathrm{exp}\left(\frac{g\mu_{B}B_{z}m}{k_{B}T}\right)g\mu_{B}B_{z}m}{\sum_{m=-J,J}\mathrm{exp}\left(\frac{g\mu_{B}B_{z}m}{k_{B}T}\right)}.
\end{equation}
Looking at the numerator and denominator, we can get the relationship of the mean magnetization to the partition function ($Z_{a}$)
\begin{equation}
\overline{\mu_{z}} = \frac{k_{B}T}{Z_{a}}\frac{\partial Z_{a}}{\partial B_{z}} = k_{B}T\frac{\partial \left(\mathrm{ln}Z_{a}\right)}{\partial B_{z}} \label{meanmagmoment}
\end{equation}
where,
\begin{equation}
Z_{a} = \sum_{m=-J,J}\mathrm{exp}\left(m\eta\right).
\end{equation}
Here we have set $\eta = g\mu_{B}B_{z}/k_{B}T$. This ratio is a parameter that links the thermal energy ($k_{B}T$), which is trying to randomly orient to atom's spin, to the external magnetic field that the paramagnetic atom tries to align with. The partition function can now be written as a geometric series and summed to give,
\begin{equation}
Z_{a} = \mathrm{e}^{\eta J}\sum_{k=0,2J}\left(\mathrm{e}^{\eta}\right)^{k} = \mathrm{e}^{\eta J} \left[ \frac{1-(\mathrm{e}^{\eta})^{2J+1}}{1-\mathrm{e}^{\eta}}\right].
\end{equation}
We then can multiply the numerator and denominator by $\exp(-\eta/2)$ and use the definition of relationship between the $\sinh$ and the exponential function to get,
\begin{equation}
Z_{a} = \frac{\sinh(J+\frac{1}{2})\eta}{\sinh(\frac{\eta}{2})}.
\end{equation}
Now we can plug our partition function back into Eqn. (\ref{meanmagmoment}) to get,
\begin{equation}
\overline{\mu_{z}} = g \mu_{B}\left[ \frac{(J+1/2)\cosh[(J+1/2)\eta]}{\sinh[(J+1/2)\eta]}-\frac{1}{2}\frac{\cosh(\eta/2)}{\sinh(\eta/2)}\right].
\end{equation}
Using that $\cosh/\sinh = \coth$ and some factoring, we will get the Brillouin function for the magnetic field,
\begin{align}
B_{J}(C) = \frac{2J+1}{2J}\coth\left(\frac{2J+1}{2J}C \right) - \frac{1}{2J}\coth\left(\frac{1}{2J}C \right),
\end{align}
where $C=J\eta$.

We could use the $\coth$ functions to calculate the magnetic field, but it would be easier to simplify the magnetic field function to get rid of the $\coth$. We can do this by considering the taylor expansion of $\coth$ for small $\eta$,
\begin{equation}
\coth{{\eta}} = \frac{1}{\eta} +\frac{1}{3}\eta + \mathcal{O}(z^{3}).
\end{equation}
We consider small $\eta$ as for most systems (including $4H$-SiC) this will be true. Using this expansion, we get
\begin{align}
&B_{J}(\eta) = \\\nonumber &\frac{1}{J}\left[ (J+\frac{1}{2})\left[\frac{1}{(J+1/2)\eta}+\frac{1}{3}(J+\frac{1}{2}\eta)\right] -\frac{1}{2}\left(\frac{2}{\eta}+\frac{\eta}{6}\right)\right].
\end{align}
This can then be reduced to give us,
\begin{equation}
B_{J}(\eta) = \left(\frac{J+1}{3}\right)\eta = \left(\frac{J+1}{3}\right)\frac{g\mu_{B}B_{z}}{k_{B}T}. \label{appenBrillField}
\end{equation}

\subsection{Larmor magnetic field} \label{append:LarmorBField}
As the electron hops from site to site in the defect center, another temperature dependent fluctuation of the magnetic field will come from Larmor precession when interacting with the new site's local magnetic environment. As the electron hops to a new nuclear site with some momentum, it will experience a Lorentz force due to the new magnetic environment and can be described by
\begin{align}
\frac{\partial \textbf{p}}{\partial t} = q\left(\textbf{v}_{q} \times \textbf{B}\right),
\end{align}
where $\textbf{p}$ is the momentum and $\textbf{v}_{q}$ is the velocity of the charge. We neglect the electric field Lorentz force as we are currently only worried about the magnetic field fluctuations of the local environment. We consider again that the nuclear magnetic field is along the z-axis allowing us to describe the velocity of the charge ($v_{q} = \sqrt{v_{x}^{2}+v_{y}^{2}+v_{z}^{2}}$) in the x-,y-, and z- directions as
\begin{align}
v_{z} &= v_{0,z} \\ \nonumber
v_{x} &= v_{0}\cos(\omega_{L}t+\phi) \\ \nonumber
v_{y} &= -v_{0}\sin(\omega_{L}t+\phi).
\end{align}
Here $\omega_{L} = ec^{2}\mathrm{B}/E_{q}$ is the Larmor frequency, $v_{0}$ and $v_{0,z}$ are components of the initial velocity, and $\phi$ is some initial phase which for simplicity we will set to zero. $E_{q}$ is the electron energy in electron volts, $q$ is the electron charge, and $c$ is the speed of light. We can now use kinematic equations to solve for the Larmor radius (which here will also be defined by the atomic radius)
\begin{equation}
 r = \frac{v_{q}E_{q}}{ec^{2}|\mathrm{B}|}.
\end{equation}
It is straight forward from here to then solve for $|\mathrm{B}|$ which we will define as $B_{L}$,
\begin{equation}
B_{L}(T) = \left(\frac{2}{qc^{2}r}\right)v_{q}(T)E_{q}(T).
\end{equation}
Here we have written this in terms of temperature dependence as both the charge velocity will be temperature dependent due to equipartition and the electron energy temperature dependence will arise from the Jahn-Teller distortion.

\subsection{Nuclear magnetic field \label{Appdx:nuclearmagfield}}
With Eqn. (\ref{appenBrillField}), we have the external magnetic field felt by the electron due to magnetic and thermal energy. There is one more piece we need to determine to be able to describe the magnetic field fully, $B_{z}$ from the nuclei the electron is bound to. If we consider that the nuclei the electron will see is a magnetic dipole, we can write the magnetic field as
\begin{equation}
B_{N} = \frac{\mu_{0}\mu_{N}}{2\pi r^{3}}. \label{appendixNucBField}
\end{equation}
Here $\mu_{0}$ is the permeability of free space, $\mu_{N}$ is the nuclear magnetic moment, and $r$ is the distance from the nuclei (or the atomic radius). From here, we need to determine the nuclear magnetic moment of the vacancy centers in $4H$-SiC. Observational and theoretical studies by Kukushinkin \textit{et al} \cite{KO22} saw that in the silicon vacancy center, $\mu_{N}$ was on the order of $3\mu_{B}$. As for the positvely and negatively charged carbon vacancies, we have found no direct study on the strength of the nuclear magnetic moment. With that, we will assume a similar strength to the silicon vacancy. With that we can write Eqn. (\ref{appendixNucBField}) as,
\begin{equation}
B_{N} = \frac{3\mu_{0}\mu_{B}}{2\pi r^{3}}
\end{equation}

\section{Holstein Two-Site Hopping}
\subsection{Holstein Hamiltonian derivation} \label{append:HolstModel}
Due to the polaronic nature of the vacancy centers in $4H$-SiC and the fact that the electron hopping will only occur between two sites (i.e., between the sites Si$_{1}-$Si$_{2}$, Si$_{1}-$Si$_{3}$, Si$_{1}-$Si$_{4}$ for the carbon vacancy or C$_{1}-$C$_{2}$, C$_{1}-$C$_{3}$, C$_{1}-$C$_{4}$ for the silicon vacancy), we can describe the hopping with Holstein's two-site hopping model \cite{TH781,TH782}. From here, we will follow the derivation done by Holstein applied to the vacancy centers in $4H$-SiC. To start, we need to consider the basic Hamiltonian of the molecular crystal,
\begin{equation}
\hat{H} = \hat{H}_{L} + \hat{H}_{e} + \hat{H}_{eL}. \label{molecularHamil}
\end{equation}
Here $\hat{H}_{L}$ is the vibrational Hamiltonian, consisting of the kinetic and potential energies of vibration of an array of independent diatomic molecules that takes into account the coupling to nearest-neighbors,
\begin{equation}
\hat{H}_{L} = \sum_{n=1}^{N}\frac{-\hbar^{2}}{2M}\frac{\partial^{2}}{\partial r_{n}^{2}} + \frac{1}{2}M\omega_{0}^{2}r_{n}^{2} + \frac{1}{2}M\omega_{1}^{2}r_{n}r_{n+1}.
\end{equation}
Here $M$ is the mass, $r_{n,n+1}$ are the positions of site $n$ and the nearest-neighbor site $n+1$, and $\omega_{0,1}$ are the vibrational frequencies. $\hat{H}_{e}$ and $\hat{H}_{eL}$ are the electron transfer and electron interaction with the local vibrational mode Hamiltonians which make up the Holstein Hamiltonian such that $\hat{H}_{e}+\hat{H}_{eL}=\hat{H}_{Hol}$ and is described by \cite{TH59,TH781,TH782,YC14,FT07},
\begin{align}
\hat{H}_{Hol} = -A\sum_{i}\sum_{j}c_{i}^{\dagger}c_{i}+\hbar \nonumber & \omega_{i}\sum_{i}b_{i}^{\dagger}b_{i}\\
&-\lambda\sum_{i}\left( b_{i}^{\dagger}+b_{i}\right)c_{i}^{\dagger}c_{i}.
\end{align}
Here $A$ is the integral amplitude of intersite hopping (i.e., the closest neighbor electron transfer parameter), $c_{i}^{\dagger}$ and $(c_{i})$ are the electron creation and annihilation operators, $b_{i}^{\dagger}(b_{i})$ are the phonon creation and annihilation operators due to interactions with vibrational modes, $\lambda$ is thus the electron-phonon coupling, and $\omega_{i}$ is the angular frequency of the phonon.

Since we are focusing on two-site hopping, let us consider sites $p$ and $p+1$ and work out Eqn. (\ref{molecularHamil}) as
\begin{align}
\hat{H} = \frac{-\hbar^{2}}{2M} & \nonumber \left( \frac{\partial^{2}}{\partial r_{p}^{2}} + \frac{\partial^{2}}{\partial r_{p+1}^{2}}\right) \\& \nonumber + \frac{1}{2}M\omega_{0}^{2}\left(r_{p}^{2} + r_{p+1}^{2}\right) + \frac{1}{2}M\omega_{1}^{2}r_{p}r_{p+1} \\ & \nonumber-\frac{1}{2}A(n^{c}_{p}+n^{c}_{p+1}) + \frac{1}{2}\Delta\left(n^{b}_{p} - n^{b}_{p+1}\right) \\ &- \lambda\left( b_{p+1}^{\dagger}c_{p} + c_{p}^{\dagger}b_{p+1}\right) + c.c..
\end{align}
Here we use $n^{c}_{p}=c^{\dagger}_{p}c_{p}$ and $n^{b}_{p}=b^{\dagger}_{p}b_{p}$ which relate to the eigenvalues of the creation and annihilation operators, $\Delta = \hbar(\omega_{p}-\omega_{p+1})$ is the energy difference and $c.c.$ represents the complex conjugate. Next, let us do a transformation to relative coordinates,
\begin{align}
&R = (r_{p}+r_{p+1})/2, \\ \nonumber
&r = r_{p+1} - r_{p}.
\end{align}
Our next step to isolate the hopping will be to split our Hamiltonian into two pieces. One piece will take into account the center-of-mass coordinate ($R$) and the other will take into account the relative coordinate piece ($r$). We then will have,
\begin{equation}
\hat{H} = \hat{H}_{CM} + \hat{H}_{rel},
\end{equation}
where the center-of-mass Hamiltonian ($H_{CM}$) is given by
\begin{equation}
\hat{H}_{CM} = \frac{-\hbar^{2}}{4M}\frac{\partial^{2}}{\partial R^{2}} + M\omega_{+}^{2}R^{2} - A(n^{c}_{p}+n^{c}_{p+1}),
\end{equation}
and the relative Hamiltonian ($\hat{H}_{rel}$) is given by
\begin{align}
\hat{H}_{rel} = & \nonumber  \frac{-\hbar^{2}}{M}\frac{\partial^{2}}{\partial r^{2}} + \frac{M}{4}\omega_{-}^{2}r^{2} - \frac{1}{2}A(n^{c}_{p}-n^{c}_{p+1}) \\ &- \lambda\left( b_{p+1}^{\dagger}c_{p} + c_{p}^{\dagger}b_{p+1}\right) - \frac{1}{2}\Delta\left(n^{b}_{p+1} - n^{b}_{p}\right), \label{appendixRelHam}
\end{align}
where $\omega_{\pm}^{2}=(\omega^{2}_{0}\pm\omega^{2}_{1})/2$. At this point, the $\hat{H}_{CM}$ is decoupled from the electron transfer. One can also see if we set $\lambda=0$ (i.e., no electron transfer), the solutions will come out to the harmonic oscillator as expected for the Holstein two-site model.

\subsection{Holstein two-site hopping rate derivation} \label{append:TwoSiteHopping}
From here, we can start with the next major step of the Holstein model and determine the two-site hopping rate which will allow us to determine the characteristic lifetime, $\tau_{c}$, of the polaron hopping. We again follow the work of Holstein and will start with the Holstein expression for the two-site hopping rate from site $p\xrightarrow{} p+1$ \cite{TH59,TH781,TH782,YC14,FT07},
\begin{equation}
\nu_{p \xrightarrow{} p+1} = \int \frac{\Phi(E_{a})}{2\pi\hbar Z_{0}} \mathrm{e}^{\frac{-E_{a}}{k_{B}T}} dE. \label{appendTwoSiteHopping}
\end{equation}
Here $\Phi(E_{a})$ is a thermodynamic average of the electron transition wave function, and $Z_{0}$ is the vibrational partition function
\begin{equation}
Z_{0} = \mathrm{e}^{\frac{-E_{ZP}}{k_{B}T}}\Pi_{i}\frac{1}{1-\mathrm{e}^{\frac{-\hbar\omega_{i}}{k_{B}T}}},
\end{equation}
where $E_{ZP}$ is the zero-point energy. As justified by Holstein \textit{et al}, the wave function can be written as \cite{TH781,TH782},
\begin{equation}
\Phi(E_{a}) = \frac{2\pi\lambda^{2}}{\hbar}\left| \frac{dV_{p}}{dr}-\frac{dV_{p+1}}{dr}\right|_{r_{0}}\mathrm{e}^{\left[ -\frac{2}{\hbar}\int^{r_{f}}_{r_{0}}P(E_{a},r)dr\right]}.
\end{equation}
Here $V_{p,p+1}$ are the potential energy piece of the Hamiltonian, $P(E_{a},r)$ is the probability factor of the electron to hop and is given by 
\begin{equation}
P(E_{a},r)=\left[M(V(r)-E_{a})\right]^{1/2},
\end{equation}
and $r_{0,f}$ are the initial and final positions of the electron. Following Holstein, we can get a great approximation for the exponent by considering the saddle-point technique for probability distribution functions and define \cite{TH781,HED54},
\begin{equation}
F(E_{a}) = \frac{E_{a}}{k_{B}T} +\left( \frac{2}{\hbar}\right)\int^{r_{f}}_{r_{0}} \left[M(V(r)-E_{a})\right]^{1/2}dr.
\end{equation}
Plugging in the potential energy part from Eqn. (\ref{appendixRelHam}), we can plug this integral back into Eqn. (\ref{appendTwoSiteHopping}) and solve for the two-site hopping rate,
\begin{equation}
\nu_{p \xrightarrow{} p+1} = \frac{1}{\tau_{c}} = \frac{\lambda^{2}}{2\hbar} \sqrt{\frac{\pi}{E_{a}k_{B}T}} e^{-\frac{E_{a}}{k_{B}T}}.
\end{equation}

\section{The Jahn-Teller Effect} \label{append:JTEffect}
The Jahn-Teller effect arises from vibronic interactions within a molecular structure. Vibronic interactions at their core are electron-nuclear interactions due to nuclear motions in both degenerate and non-degenerate states. To derive the necessary constants we need, we will follow the textbook of Bersuker \cite{IBB06}. The Hamiltonian for the vibronic interactions is generally given by
\begin{align}
    H = H_{r} + H_{Q} + V(r,R),
\end{align}
where $H_{r}$ is the electronic component which includes the kinetic energy of the electrons and the interelectronic electrostatic interaction, $H_{Q}$ is the kinetic energy of the nuclei, and $V(r,Q)$ is the interaction potential energy operator between the nuclei and electrons.

To describe the vibronics of the molecular system, we are most interested in the operator $V(r,Q)$. The potential energy operator is commonly defined as,
\begin{equation}
V(r,R) = \sum_{i,\alpha}h_{i,\alpha} + \sum_{\alpha\beta}G_{\alpha,\beta},
\end{equation}
where
\begin{equation}
h_{i,\alpha} = \frac{-e^{2}Z_{\alpha}}{\left| \textbf{r}_{i}-\textbf{R}_{\alpha}\right|},
\end{equation}
and
\begin{equation}
G_{\alpha,\beta} = \frac{-e^{2}Z_{\alpha}Z_{\beta}}{\left| \textbf{R}_{\alpha}-\textbf{R}_{\beta}\right|}
\end{equation}
Here $\textbf{r}_{i}$ is the position of the electrons and ${R}_{\alpha,\beta}$ are the positions of nuclei. If we neglect nuclear displacements, we get the Born-Oppenheimer approximation. Instead, let us consider the expansion of small displacements of the operator about a symmetrized point $Q$. The expansion then gives us,
\begin{align}
    V(r,Q) = V(r,0) + \nonumber &\sum_{\alpha} \left(\frac{\partial V}{\partial Q_{\alpha}}\right) Q_{\alpha}\\ 
    & + \frac{1}{2}\sum_{\alpha,\beta}\left( \frac{\partial^{2} V}{\partial Q_{\alpha} \partial Q_{\beta}} \right) Q_{\alpha}Q_{\beta} + \cdots, \label{appendPotenOper}
\end{align}
The first term in Eqn. (\ref{appendPotenOper}) is the electrons in the field of the nuclei at $Q=0$. From this, one can solve the Schr\"{o}dinger equation
\begin{equation}
\left[H_{r} +V_{r,0} - \epsilon^{'}_{k}\right]\varphi_{k}(r)=0.
\end{equation}
Here $\epsilon^{'}_{k}$ is a set of eigenenergies and $\varphi_{k}(r)$ are a set of wavefunctions for a given nuclear configuration corresponding to the point $Q=0$. To solve the full Schr\"{o}dinger equation, we can expand the wavefunctions to consider nuclear displacements,
\begin{equation}
\Phi(r,Q) = \sum_{k}\chi_{k}(Q)\varphi_{k}(r),
\end{equation}
where $\chi_{k}(Q)$ are the expansion coefficients that are functions of $Q$. We can now plug these wavefunctions back into the Schr\"{o}dinger equation to get,
\begin{equation}
\left[H_{Q} + \epsilon_{k}(Q) - E\right]\chi_{k}(Q) +\sum_{m\neq k}W_{km}(Q)\chi_{m}(Q)=0.
\end{equation}
Here $W_{km}(Q)$ denotes the electronic matrix element of the vibronic interactions that comes from the $Q$ dependent part of $V(r,Q)$. From this, we can define a potential energy as,
\begin{align}
W(r,Q) &= V(r,Q) - V(r,0) \\ \nonumber
&= \sum_{\alpha}\left( \frac{\partial V}{\partial Q_{\alpha}}\right)Q_{\alpha} \\ \nonumber &+ \frac{1}{2}\sum_{\alpha,\beta}\left( \frac{\partial^{2}}{\partial Q_{\alpha}\partial Q_{\beta}}\right)Q_{\alpha}Q_{\beta} + \cdots .
\end{align}
From here, we can define different coupling constants which will describe the measure of influence of the nuclear displacements on the electronic distribution, as well as, the effect upon the nuclear dynamics by the changing electron structure. This is one of the key insights that can be gained from the Jahn-Teller effect as these changes in electronic structure and distribution can be spontaneous, breaking the local symmetry. We can define these coupling constants to be \cite{IBB06,ENG11}
\begin{align}
    F &= \langle \Phi_{n}| \left( \frac{\partial V}{\partial Q_{\alpha}}\right) |\Phi_{n^{'}}\rangle, \nonumber \\
    G &= \langle \Phi_{n}| \left( \frac{\partial^{2} V}{\partial Q_{\alpha} \partial Q_{\beta}} \right)_{\alpha \neq \beta} |\Phi_{n^{'}}\rangle, \nonumber \\
    K &= \langle \Phi_{n}| \left( \frac{\partial^{2} V}{\partial Q_{\alpha} \partial Q_{\beta}} \right)_{\alpha = \beta} |\Phi_{n^{'}}\rangle.\label{appendcoupling}
\end{align}
where $F$ is the linear vibronic coupling constant, $G$ is the quadratic vibronic coupling constant, and $K$ is the force constant of the nuclear vibrations.

Lastly, we are interested in the Jahn-Teller energy ($E_{JT}$) that depends on these vibronic couplings. To determine $E_{JT}$, we can calculate the adiabatic potential energy surface (APES). To do this, let us consider what the elements, $W_{km}$, will be for a single $k$ and $m$. We consider a single $k$ and $m$ as these will dictate two separate symmetry groups as shown by Bersuker \cite{IBB06}. The matrix of elements $W_{km}$ using our definitions for the coupling constants will be,
\begin{equation}
W = F\left( Q_{k}\hat{\sigma}_{z} - Q_{m}\hat{\sigma}_{x}\right) + G\left[\left(Q_{k}^{2}-Q_{m}^{2} \right)\hat{\sigma}_{z} + 2Q_{k}Q_{m}\hat{\sigma}_{x}\right],
\end{equation}
where $\hat{\sigma}_{z,x}$ are the Pauli matrices. If we now consider the eigenenergies of the system which take into account these vibronic couplings we will have in symmetrized coordinates \cite{IBB06},
\begin{equation}
\epsilon = \frac{1}{2}K\rho^{2} \pm \rho\left[ F^{2}+G^{2}\rho^{2}+2FG\rho\cos3\phi\right]^{1/2}.
\end{equation}
Here $\rho=\sqrt{(Q_{x}^{2}+Q_{y}^{2})}$ and $\phi=\arctan(Q_{y}/Q_{x})$.

\end{document}